\documentclass[12pt,english]{article}

\usepackage[latin9]{inputenc}
\usepackage[letterpaper]{geometry}
\usepackage{units}
\usepackage{textcomp}
\usepackage{amsmath}
\usepackage{amssymb}
\usepackage{graphicx}
\usepackage{setspace}
\usepackage{esint}

\makeatletter

\newcommand{\lyxmathsym}[1]{\ifmmode\begingroup\def\b@ld{bold}
  \text{\ifx\math@version\b@ld\bfseries\fi#1}\endgroup\else#1\fi}

\providecommand{\tabularnewline}{\\}

\newenvironment{lyxlist}[1]
{\begin{list}{}
{\settowidth{\labelwidth}{#1}
 \setlength{\leftmargin}{\labelwidth}
 \addtolength{\leftmargin}{\labelsep}
 }}
{\end{list}}

\@ifundefined{date}{}{\date{}}
\usepackage{graphicx}
\usepackage{indentfirst}
\usepackage{slashed}

\makeatother

\usepackage{babel}
\begin{document}
\noindent \begin{center}
\thispagestyle{empty}
\par\end{center}

\begin{center}
RADIATIVE SYMMETRY BREAKING IN THE 
\par\end{center}

\begin{center}
SUPERSYMMETRIC MINIMAL $B-L$ 
\par\end{center}

\begin{center}
EXTENDED STANDARD MODEL\vspace*{0.1in}

\par\end{center}

\begin{singlespace}
\begin{center}
by
\par\end{center}

\begin{center}
ZACHARY BURELL
\par\end{center}

\begin{center}
NOBUCHIKA OKADA, COMMITTEE CHAIR
\par\end{center}
\end{singlespace}

\begin{center}
LOUIS J. CLAVELLI\\
CONOR HENDERSON\\
GREG SZULCZEWSKI\vspace*{0.27in}

\par\end{center}

\noindent \begin{center}
A THESIS
\par\end{center}

\vspace*{0.29in}

\begin{singlespace}
\noindent \begin{center}
Submitted in partial fulfillment of the requirements\\
for the degree of Master of Science in the \\
Department of Physics and Astronomy \\
in the Graduate School of \\
The University of Alabama
\par\end{center}
\end{singlespace}

\begin{doublespace}
\noindent \begin{center}
\vspace*{0.57in}
TUSCALOOSA, ALABAMA
\par\end{center}

\noindent \begin{center}
\vspace*{0.2in}
2011
\par\end{center}
\end{doublespace}

\noindent \begin{center}
\pagenumbering{roman}
\par\end{center}

\noindent \begin{center}
\clearpage{}
\par\end{center}

\noindent \begin{center}
\vspace*{0.75in}
ABSTRACT
\par\end{center}

\begin{doublespace}
The Standard Model (SM) of particle physics is a precise model of
electroweak interactions, however there is growing tension between the SM and observations (neutrino
oscillations, dark matter, dark energy, baryogenesis, among others).  There is no reason to expect the validity of the ad hoc SM to remain intact at energy scales above a few TeV, thus a more fundamental theory will almost certainly
be required. 

Motivated by these considerations, we investigate a Supersymmetric version of a natural extension of the SM, the $U(1)_{B-L}$ model, that is obtained by gauging the accidental B-L symmetry that exists in the ordinary SM. The Supersymmetric $U(1)_{B-L}$ extended SM can resolve the neutrino mass problem, the dark matter problem, the hierarchy problem, and provides a mechanism for establishing the observed baryon asymmetry of the Universe. 

When we include quantum corrections to the Higgs potential of the model, we find that Radiative $B-L$ symmetry
breaking occurs through the interplay between large Majorana Yukawa
couplings and SUSY breaking masses, and the SM neutrino masses arise through the Type-I seesaw mechanism. We deduce  from the form RGEs that $B-L$ will be broken near the TeV scale when the Majorana Yukawa couplings are of order 0.5, leading to TeV scale right-handed neutrinos, which could be accessible at the LHC. We discuss the different vacua of the theory and the viability of a right-handed neutrino  dark matter candidate in the R-parity conserving and violating sectors. A new $\mathbb{Z}_{2}$ parity is postulated in the R-parity violating sector in which case the lightest right handed neutrino becomes a viable Dark Matter candidate in addition to the gravitino.

\end{doublespace}

\noindent \begin{center}
\clearpage{}
\par\end{center}

\vspace*{1.35in}

\noindent \begin{center}
DEDICATION
\par\end{center}

\begin{doublespace}
My thesis is dedicated to my family whose love and support made writing
this thesis a possibility.
\end{doublespace}

\clearpage{}

\noindent \begin{center}
LIST OF ABBREVIATIONS AND SYMBOLS\vspace*{0.1in}

\par\end{center}
\begin{lyxlist}{00.00.0000}
\item [{SM}] the Standard Model 
\item [{GeV}] $10^{9}$ electron-volts
\item [{TeV}] $10^{12}$ electron-volts
\item [{EWSB}] Electroweak symmetry breaking
\item [{VEV}] Vacuum expectation value
\item [{GUT}] Grand unified theory
\item [{SSB}] Spontaneous symmetry breaking
\item [{SUSY}] Supersymmetry/Supersymmetric 
\item [{MSSM}] Minimal Supersymmetric Standard Model
\item [{LHC}] Large Hadron Collider
\item [{mSUGRA}] Minimal Supergravity
\item [{FCNC}] Flavor changing neutral currents
\item [{$x^{\mu}$}] Space time coordinates $\left(\mu=0,1,2,3\right)$
\item [{$g^{\mu\nu}$}] Metric tensor $\mathrm{diag}\left(1,-1,-1,-1\right)$
\item [{$g_{s}$}] Strong coupling constant
\item [{$g$}] Weak coupling constant
\item [{$g'$}] $U\left(1\right)_{\mathrm{Y}}$ coupling constant
\item [{$g_{BL}$}] The $U\left(1\right)_{B-L}$ coupling constant.
\item [{$\mathcal{L}$}] Lagrangian density
\item [{$\Phi$}] Left handed chiral superfield
\item [{$\phi$}] Left handed scalar component of chiral superfield
\item [{$\theta_{\alpha},\overline{\theta}_{\dot{\alpha}}$}] Anti-commuting
grassmanian coordinates of chiral superspace $\left(\alpha,\dot{\alpha}=1,2\right)$
\item [{$\nu_{i}^{c}$}] Right handed neutrinos$\left(i=1,2,3\right)$
\item [{$e^{c}$}] Right handed positrons
\item [{$\mathrm{Y}$}] Yukawa coupling
\item [{$H_{u}$}] MSSM {}``up'' type Higgs left handed chiral superfield
\item [{$H_{d}$}] MSSM {}``down'' type Higgs left handed chiral superfield
\item [{$\mathcal{W}$}] Super-potential
\item [{$\mathcal{K}$}] Kahler-potential
\item [{$M_{GUT}$}] Grand unification scale $\left(M_{GUT}\sim10^{16}\,\mathrm{GeV}\right)$
\item [{$M_{PL}$}] Planck scale-the fundamental scale of nature $\left(M_{PL}\sim10^{19}\,\mathrm{GeV}\right)$
\item [{$\Lambda_{S}$}] Supersymmetry breaking scale
\item [{$B$}] Baryon number
\item [{$L$}] Lepton number
\item [{$B-L$}] Baryon number - Lepton number
\end{lyxlist}
\clearpage{}

\noindent \begin{center}
\vspace*{0.75in}
ACKNOWLEDGMENTS\vspace*{0.1in}

\par\end{center}

\begin{doublespace}
I would like to thank Lou Clavelli for his encouragement and guidance
and for many helpful discussions which helped me organize my knowledge.
I would thank Nobu Okada for the many lectures and hours spent guiding
me through the details of Renormalization procedures and grand unified
models. 

I would also like to thank the members of my committee for taking
the time out of their schedule to participate in the process. Last
but certainly not least I would like to thank Steve R. Best for his
guidance and for his commitment to scientific excellence which he
instilled in me during my undergraduate years working for him at the
Space Research Institute at Auburn University.
\end{doublespace}

\clearpage{}

\noindent \begin{center}
CONTENTS
\par\end{center}

\vspace*{0.1in}

\begin{doublespace}
\noindent ABSTRACT. . . . . . . . . . . . . . . . . . . . . . . .
. . . . . . . . . . . . . . . . . . . . . . . .ii

\noindent DEDICATION. . . . . . . . . . . . . . . . . . . . . . .
. . . . . . . . . . . . . . . . . . . . . . .iii

\noindent LIST OF ABBREVIATIONS AND SYMBOLS. . . . . . . . . . . .
. . . . . . . . . . . . . . .iv

\noindent ACKNOWLEDGMENTS. . . . . . . . . . . . . . . . . . . . .
. . . . . . . . . . . . . . . . . . .vi

\noindent LIST OF TABLES . . . . . . . . . . . . . . . . . . . . .
. . . . . . . . . . . . . . . . . . . . . viii

\noindent LIST OF FIGURES . . . . . . . . . . . . . . . . . . . .
. . . . . . . . . . . . . . . . . . . . .viiii

\noindent 1. THE STANDARD ELECTROWEAK MODEL. . . . . . . . . . . .
. . . . . . . . . . . . .1

\noindent 2. SUPERSPACE AND SUPERSYMMETRY . . . . . . . . . . . .
. . . . . . . . . . . . . . .18

\noindent 3. THE \emph{B-L} EXTENSION OF THE MSSM. . . . . . . . .
. . . . . . . . . . . . . . . . . . . 35

\noindent REFERENCES. . . . . . . . . . . . . . . . . . . . . . .
. . . . . . . . . . . . . . . . . . . . . . 46
\end{doublespace}

\begin{doublespace}
\noindent \begin{center}
\clearpage{}
\par\end{center}
\end{doublespace}

\noindent \begin{center}
LIST OF TABLES\vspace*{0.1in}

\par\end{center}
\begin{enumerate}
\begin{doublespace}
\item \noindent The Gell Mann Matrices . . . . . . . . . . . . . . . . .
. . . . . . . . . . . . . . . . . . . .4
\item \noindent The Standard Model Fermion Hypercharges. . . . . . . . .
. . . . . . . . . . . . . . . .7
\item \noindent The particle contents of the MSSM. . . . . . . . . . . .
. . . . . . . . . . . . . . . . . . 26
\item \noindent The particle contents of the $B-L$ extended MSSM. . . .
. . . . . . . . . . . . . . . .36
\item \noindent The particle contents of the $B-L$ extended MSSM with $\mathbb{Z}_{2}$
parity. . . . . . . . . . .45
\end{doublespace}

\end{enumerate}
\begin{center}
\clearpage{}
\par\end{center}

\noindent \begin{center}
LIST OF FIGURES\vspace*{0.1in}

\par\end{center}

\begin{doublespace}

1. One-loop quantum corrections to the Higgs mass square parameter.
. . . . . . . . .19

2. RGE running of the scalar masses in $R$-parity conserving $B-L$ model
  . . . . . . .42

3. RGE running of the scalar masses in $R$-parity violating
 $B-L$ model. . . . . . . . 43

4. RGE running of the scalar masses in the dormant case of the $B-L$
model. . . . .44

\noindent \begin{center}
\clearpage{}
\par\end{center}
\end{doublespace}

\begin{spacing}{0.90000000000000002}
\noindent \begin{center}
\pagenumbering{arabic}
\par\end{center}
\end{spacing}

\section*{Chapter 1 }

\section*{The Standard Electroweak Model}

\subsection*{1.1 {\Large Introduction}}

\begin{doublespace}
The Standard Model is certainly a triumphant stepping stone in the
progression to elucidate nature's fundamental properties; it accurately
describes the fundamental particles, the quarks and leptons, and their
interactions, with the exception of gravity, down to distance scales
$d\sim10^{-16}\,\mathrm{cm}$ \cite{Langacker2009,Abazov2006}.
\end{doublespace}

\begin{doublespace}
The SM is not flawless however, and many observations cannot be explained
by it alone. These include solar and atmospheric neutrino oscillations
\cite{Cleveland1998,Maltoni2004,Bandyopadhyay2003,Fogli2006,Ashie2005,Fukuda2002},
the existence of non-baryonic dark matter \cite{Bennett2003,Spergel2003,Spergel2007,Clowe2006},
the observed baryon asymmetry of the Universe \cite{Feng2010}, the
Muon anomalous magnetic moment \cite{Amsler2008}, and the recent
anomalous like sign di-muon charge asymmetry measured by the D$\slashed{0}$
experiment \cite{Abazov2010}. Moreover, there are a number of different
theoretical suggestions for the incompleteness of the SM, the most
pressing among these is the gauge hierarchy problem, which we discuss
in more detail in section 1.2.
\end{doublespace}

\begin{doublespace}
Neutrino flavor oscillations cannot be understood without an extension
of the SM, because the neutrino is massless in the standard framework.
However, the phenomenon of neutrino oscillations arises only in the
case of distinct non-zero neutrino masses between different neutrino
flavors. The unambiguous observation of neutrino oscillations has
presented a challenge to theorists for decades and formally speaking,
it remains a challenge. However, the introduction of one or more additional
heavy right handed neutrinos into the model gives rise to the see-saw
mechanism, which can explain the smallness of the neutrino masses
under certain additional requirements.

The existence of non-baryonic dark matter is another clear indication
that an extension of the SM is required. Dark matter is not composed
of quarks and leptons, and thus does not fall within the framework
of ordinary matter, i.e. the SM. Evidence for dark matter has been
accumulating now for about three quarters of a century, beginning
with Fritz Zwicky's observation in 1933 that galaxies in the coma
cluster were spinning much faster than could be accounted for by the
observed amount of normal matter they contained \cite{Zwicky1933}.
The measurements of individual galaxy rotation curves performed by
Vera Rubin and Albert Bosma in 1970's, confirmed Zwicky's observations,
also suggesting the existence of dark matter \cite{Bosma1978,Rubin1980}.
These classical observations have been confirmed and extended by more
recent, higher precision observations. Data from weak \cite{Refregier2003},
and strong \cite{Tyson1998} gravitational lensing, the Bullet Cluster
\cite{Clowe2006}, Big Bang Nucleosynthesis \cite{Fields2008}, Large
scale structure \cite{Allen2003}, distant supernovae \cite{Riess1998},
and the cosmic microwave background (CMB) radiation \cite{Komatsu2011},
are consistent and all suggest that Dark Matter exists in five times
the abundance of the ordinary matter explained by the SM. The incompleteness
of the SM is evident from its inability to account for these observations
without extension or modification. 
\end{doublespace}

\subsection*{1.2 The Standard Model}

\begin{doublespace}
The Standard Model is a gauge field theory of three generations of
matter based on the gauge group $\,\mathcal{G}_{SM}=SU\left(3\right)_{c}\otimes SU\left(2\right)_{L}\otimes U\left(1\right)_{Y}\,$,
each with its own running coupling constant, $\, g_{s},\: g,\: g'$.
We first present the bare SM formalism, and then introduce spontaneous
symmetry breaking (SSB). We conclude with the massive SM Lagrangian
after SSB. 

For the SM, the Lagrangian is 
\begin{equation}
\mathcal{L}_{SM}=\mathcal{L}_{SU\left(3\right)}+\mathcal{L}_{SU\left(2\right)\otimes U\left(1\right)}
\end{equation}
 where $\,\mathcal{L}_{SU\left(3\right)}\,$ is the perturbative QCD
sector Lagrangian and $\,\mathcal{L}_{SU\left(2\right)\otimes U\left(1\right)}\,$
is the Lagrangian for the electroweak sector. 
\end{doublespace}

\subsection*{1.3 The QCD sector}

\begin{doublespace}
The strong interaction arises from color charge. The fermions which
carry color quantum numbers are the quarks. There is an associated
set of massless gauge bosons, the \emph{gluons}, which also carry
color and mediate the strong-color interaction, the theory which elucidates
this phenomenology is Quantum Chromodynamics or QCD. Yukawa was among
the first to consider {}``strong'' interactions during the 1930's,
motivated by the newly discovered pion mediated nucleon-nucleon interactions
being observed at that time. Finally in 1964 Murray Gell-Mann \cite{Gell-Mann1964}
and Georg Zweig \cite{Zweig1964} introduced the eight-fold way phenomenology
of color mediated interactions. 

The perturbative QCD Lagrangian is
\begin{equation}
\ensuremath{\mathcal{L}_{SU\left(3\right)}=-\frac{1}{4}\mathcal{F}_{\mu\nu}^{i}\mathcal{F}^{i\mu\nu}+\sum_{r}\bar{Q}_{r\alpha}\, i\,\slashed{\mathcal{D}}_{\beta}^{\alpha}\, Q_{r}^{\beta}}\label{eq:4}
\end{equation}
where we sum over quark flavor $r\in\left[\mathrm{up,\,}\mathrm{down,}\,\mathrm{charm,\,\mathrm{strange,\,\mathrm{bottom,\,}}\mathrm{top}}\right]$
and $\alpha,\beta=1,2,3$ are color indices. The non-abelian field
strength tensor for the gluon fields $\,\mathcal{G}_{\mu}^{i}\,$
is, (for $i,j,k=1,\cdots,8$) : 
\begin{equation}
\mathcal{F}_{\mu\nu}^{i}=\partial_{\mu}\mathcal{G}_{\nu}^{i}-\partial_{\nu}\mathcal{G}_{\mu}^{i}-g_{s}\, f_{ijk}\,\mathcal{G}_{\mu}^{j}\,\mathcal{G}_{v}^{k}
\end{equation}
where $g_{s}$ is the QCD gauge coupling constant. The $\mathcal{F}^{2}$
term leads to three and four point gluon self interactions.

The structure constants $\, f_{ijk}$$\left(i,j,k=1,\cdots,8\right)$
are defined by $\left[\lambda^{i},\lambda^{j}\right]=2if_{ijk}\lambda^{k}$,
where the $\,\lambda^{i}\,$ are the Gell-Mann matrices, shown in
table 1 below.
\end{doublespace}

\begin{singlespace}
\noindent \begin{center}
\begin{tabular}{|ccc|}
\hline 
 &  & \tabularnewline
$\lambda^{1}=\left(\begin{array}{ccc}
0 & 1 & 0\\
1 & 0 & 0\\
0 & 0 & 0
\end{array}\right)$ & $\lambda^{2}=\left(\begin{array}{ccc}
0 & -i & 0\\
i & 0 & 0\\
0 & 0 & 0
\end{array}\right)$ & $\lambda^{3}=\left(\begin{array}{ccc}
1 & 0 & 0\\
1 & -1 & 0\\
0 & 0 & 0
\end{array}\right)$\tabularnewline
 &  & \tabularnewline
$\lambda^{4}=\left(\begin{array}{ccc}
0 & 0 & 1\\
0 & 0 & 0\\
1 & 0 & 0
\end{array}\right)$ & $\lambda^{5}=\left(\begin{array}{ccc}
0 & 0 & -i\\
0 & 0 & 0\\
i & 0 & 0
\end{array}\right)$ & $\lambda^{6}=\left(\begin{array}{ccc}
0 & 0 & 0\\
0 & 0 & 1\\
0 & 1 & 0
\end{array}\right)$\tabularnewline
 &  & \tabularnewline
$\lambda^{7}=\left(\begin{array}{ccc}
0 & 0 & 0\\
0 & 0 & -i\\
0 & i & 0
\end{array}\right)$ & $\lambda^{8}=\frac{1}{\sqrt{3}}\left(\begin{array}{ccc}
1 & 0 & 0\\
0 & 1 & 0\\
0 & 0 & -2
\end{array}\right)$ & \tabularnewline
 &  & \tabularnewline
\hline 
\end{tabular}
\par\end{center}
\end{singlespace}

\noindent \begin{center}
Table 1. The Gell Mann Matrices
\par\end{center}

\begin{doublespace}
The second term in $\mathcal{L}_{SU\left(3\right)}$ is the gauge
covariant derivative for \emph{$SU(3)_{c}$}: 
\begin{equation}
\mathcal{D}_{\mu\beta}^{\alpha}=\left(\mathcal{D}_{\mu}\right)_{\alpha\beta}=\mathcal{\partial_{\mu}}\delta_{\alpha\beta}+ig_{s}\mathcal{G}_{\mu}^{i}L_{\alpha\beta}^{i}
\end{equation}
where the quarks transform according to the triplet representation
matrices $L^{i}=\lambda^{i}/2$. The Lagrangian $\mathcal{L}_{SU\left(3\right)}$
is invariant under the $SU\left(3\right)$ gauge transformations.

The purely vector, hence parity conserving QCD color interactions
are flavor diagonal but in general can mix quark colors. Also, it
is clear from equation \ref{eq:4} that there are no bare mass terms
for the quarks in the Lagrangian. Bare mass terms would be possible
in QCD alone but the electroweak sector of the unified model has global
chiral symmetry in the unbroken electroweak phase which forbids bare
mass terms. The quark masses come in through spontaneous symmetry
breaking. There are additional, effective ghost and gauge-fixing terms
which enter into the quantization of both $SU\left(3\right)_{C}$
and $SU\left(2\right)_{L}\otimes U\left(1\right)_{Y}$ sectors of
the theory. The QCD formalism is very extended and we will not be
using it in what follows, so we omit these terms in this overview
of the SM. 
\end{doublespace}

\begin{doublespace}

\subsection*{1.4 The Electroweak Sector}
\end{doublespace}

\begin{doublespace}
The Electroweak Lagrangian is 

\begin{equation}
\mathcal{L}_{SU\left(2\right)\otimes U\left(1\right)}=\mathcal{L}_{gauge}+\mathcal{L}_{\varphi}+\mathcal{L}_{f}+\mathcal{L}_{\mathrm{Yukawa}}
\end{equation}
The gauge component of the Lagrangian is:
\begin{equation}
\mathcal{L}_{gauge}=-\frac{1}{4}\mathcal{F}_{\mu\nu}^{i}\mathcal{F}^{i\mu\nu}-\frac{1}{4}\mathcal{B}_{\mu\nu}\mathcal{B}^{\mu\nu}
\end{equation}
where the abelian $\,\mathcal{B}_{\mu\nu}\,$ and non-abelian $\,\mathcal{F}_{\mu\nu}^{a}\,$
field strength tensors are: 
\begin{equation}
\mathcal{B}_{\mu\nu}=\mathcal{\partial}_{\mu}\mathcal{B}_{\nu}-\mathcal{\partial}_{\nu}\mathcal{B}_{\mu}
\end{equation}
\begin{equation}
\mathcal{F}_{\mu\nu}^{a}=\partial_{\mu}W_{\nu}^{a}-\partial_{\nu}W_{\mu}^{a}+g\epsilon_{abc}W_{\mu}^{b}W_{v}^{c},
\end{equation}
The superscript $\,'a\,'\,$\emph{ }runs over the adjoint representation
of the gauge group. The spacetime index $\,\mu\,$ runs from 0...3,
and the structure constant $\,\epsilon_{abc}\,$ is totally anti-symmetric;
\emph{$\:$$g$$\,$} is the \emph{$SU(2)$} gauge coupling, and\emph{
g'} is the \emph{$\, U(1)\,$} gauge coupling. The fields $\, W_{\mu}^{a}\,$
and $\,\mathcal{B}_{\mu}\,$ are the \emph{$SU(2)$} and \emph{$U(1)$}
gauge fields, respectively. After Spontaneous Symmetry Breaking (SSB),
the \emph{$\,\mathcal{B}\,$} field mixes with $\, W_{3}\,$ to form\emph{
}the neutral massless photon $\,\gamma\,$, the bosonic mediator of
the Electro-magnetic interaction, and the weak neutral massive $Z$
gauge boson, the neutral massive weak mediator.

The scalar spin-zero Higgs component of the Lagrangian is: 
\begin{equation}
\mathcal{L}_{\varphi}=\left(\mathcal{D}^{\mu}\varphi\right)^{\dagger}\left(\mathcal{D}_{\mu}\varphi\right)-V\left(\varphi\right)
\end{equation}
where, 
\begin{equation}
\varphi=\left(\begin{array}{c}
\varphi^{+}\\
\varphi^{0}
\end{array}\right)
\end{equation}
 is a complex $SU\left(2\right)$ doublet with hypercharge $\,\boldsymbol{y}_{\varphi}=1/2$.
The gauge covariant derivative is:
\begin{equation}
\mathcal{D}_{\mu}\varphi=\left(\partial_{\mu}+i\, g\frac{\tau^{i}}{2}W_{\mu}^{i}+\frac{i\, g'}{2}\mathcal{B}_{\mu}\right)\varphi
\end{equation}
 where $\,\tau^{i}\,$ are the Pauli matrices: 
\begin{equation}
\tau^{1}=\left(\begin{array}{cc}
0 & 1\\
1 & 0
\end{array}\right),\;\tau^{2}=\left(\begin{array}{cc}
0 & -i\\
i & 0
\end{array}\right),\;\tau^{3}=\left(\begin{array}{cc}
1 & 0\\
0 & -1
\end{array}\right).
\end{equation}

The scalar Higgs potential $\, V\left(\varphi\right)\,$ is given
by all self interaction terms which satisfy $\, SU\left(2\right)\otimes U\left(1\right)\,$
gauge invariance. There are only two such terms, so the scalar potential
is: 
\begin{equation}
V\left(\varphi\right)=\mu^{2}\varphi^{\dagger}\varphi+\lambda\left(\varphi^{\dagger}\varphi\right)^{2}
\end{equation}
It is relevant to point out that vacuum stability requires that $\,\lambda>0\,$,
moreover, if $\,\mu^{2}<0\,$ we get SSB. 

The electroweak theory is chiral, and parity is violated in the electroweak
sector by assigning left handed and right handed matter fields: 
\begin{equation}
\,\psi_{L}=\left(1-\gamma_{5}\right)\frac{\psi}{2}\;\quad,\quad\;\psi_{R}=\left(1+\gamma_{5}\right)\frac{\psi}{2}\,
\end{equation}
 to different representations of $\, SU\left(2\right)\otimes U\left(1\right)$.
The fermion sector consists of $\, m=3\,$ families of left handed
quark and lepton $SU\left(2\right)$ doublets:
\begin{equation}
Q_{mL}=\left(\begin{array}{c}
U_{m}\\
D_{m}
\end{array}\right)\:\quad,\quad\: L_{mL}=\left(\begin{array}{c}
\nu_{m}\\
E_{m}
\end{array}\right)
\end{equation}
with corresponding right handed singlets: $U_{mR},\; D_{mR},\; E_{mR}$.
The left handed $\,\psi_{L}\,$, and right handed $\,\psi_{R}\,$
matter fields transform under $\, U\left(1\right)\,$ in such a way
that electric charge is given by $\,\boldsymbol{q}=t^{3}+\boldsymbol{y}$,
where the $\, t^{i}\:\left(i=1,2,3\right)$ are the generators of
weak isospin and $\,\boldsymbol{y}\,$ is the hypercharge. The hypercharges
for the doublets and singlets are given below in table 2:
\end{doublespace}

\begin{singlespace}
\noindent \begin{center}
\begin{tabular}{|c|c|c|c|c|c|}
\hline 
 & $Q_{L}$ & $L_{L}$ & $U_{R}$ & $D_{R}$ & $E_{R}$\tabularnewline
\hline 
\hline 
$\boldsymbol{y}$ & $+\nicefrac{1}{6}$ & $-\nicefrac{1}{2}$ & $+\nicefrac{2}{3}$ & $-\nicefrac{1}{3}$ & $-1$\tabularnewline
\hline 
\end{tabular}
\par\end{center}
\end{singlespace}

\noindent \begin{center}
Table 2: The SM fermion hypercharges
\par\end{center}

\begin{doublespace}
The fermion component of the Lagrangian is (note that the weak eigenstates
$Q_{mL}$, $L_{mL}$ etc. are generally mixtures of the mass eigenstates.)
: 
\begin{equation}
\mathcal{L}_{f}=\sum_{m}^{F}\left(\bar{Q}_{mL}\, i\,\slashed{\mathcal{D}}Q_{mL}+\bar{L}_{mL}\, i\,\slashed{\mathcal{D}}L_{mL}+\bar{U}_{mR}\, i\,\slashed{\mathcal{D}}U_{mR}+\bar{D}_{mR}\, i\,\slashed{\mathcal{D}}D_{mR}+\bar{E}_{mR}\, i\,\slashed{\mathcal{D}}E_{mR}\right)
\end{equation}

The gauge covariant derivatives for the individual matter fields are: 
\end{doublespace}

\begin{equation}
\mathcal{D}_{\mu}Q_{mL}=\left(\partial_{\mu}+i\, g\frac{\tau^{i}}{2}\, W_{\mu}^{i}+\frac{ig'}{6}\,\mathcal{B}_{\mu}\right)Q_{mL}
\end{equation}

\begin{equation}
\mathcal{D}_{\mu}L_{mL}=\left(\partial_{\mu}+i\, g\frac{\tau^{i}}{2}\, W_{\mu}^{i}-\frac{ig'}{2}\,\mathcal{B}_{\mu}\right)L_{mL}
\end{equation}

\begin{equation}
\mathcal{D}_{\mu}U_{mR}=\left(\partial_{\mu}+i\,\frac{2g'}{3}\,\mathcal{B}_{\mu}\right)U_{mR}
\end{equation}

\begin{equation}
\mathcal{D}_{\mu}D_{mR}=\left(\partial_{\mu}-i\,\frac{g'}{3}\,\mathcal{B}_{\mu}\right)D_{mR}
\end{equation}

\begin{equation}
\mathcal{D}_{\mu}E_{mR}=\left(\partial_{\mu}-i\, g'\mathcal{B}_{\mu}\right)E_{mR}
\end{equation}

\begin{doublespace}
The chiral symmetry forbids any bare mass terms for the fermions.
We can read off interactions between the gauge fields $\, W_{\mu}^{i}\,$
and $\,\mathcal{B}_{\mu}\,$ and the fermion fields $\,\psi_{L/R}\,$
from the covariant derivatives. 

Finally, we must introduce a Yukawa component which couples the left
handed $\,\psi_{L}\,$ and right handed $\,\psi_{R}\,$ matter fields.
The Yukawa Lagrangian is: 
\begin{equation}
-\mathcal{L}_{{\rm Yukawa}}=\sum_{m,n=1}^{F}\left[Y_{mn}^{u}\bar{Q}_{mL}\varphi^{c}U_{nR}+Y_{mn}^{d}\bar{Q}_{mL}\varphi\, D_{nR}+Y_{mn}^{e}\bar{L}_{mn}\,\varphi\, E_{nR}\right]+{\rm h.c.}
\end{equation}
The matrices $\, Y_{mn}^{\left\{ \psi\right\} }\,$ will generate
mass terms after SSB, and they describe the Yukawa couplings between
the single Higgs doublet $\,\varphi\,$, and the various flavors \emph{m}
and \emph{n} of quarks and leptons. To give mass to the up quarks,
down quarks, and the electron, we need Higgs representations with
$\,\boldsymbol{y}_{\varphi}=-\frac{1}{2}\,$ and $\,\boldsymbol{y}_{\varphi}=\frac{1}{2}\,$.
However, we do not have to introduce an additional Higgs to achieve
this, since we are dealing with $\, SU(2)\,$ doublets. We only need
one Higgs in this case because the conjugate representation $\,\bar{2}\,$
of the group $\, SU(2)$, is related to the normal representation
$\,2\,$, by a similarity transformation, i.e., 
\begin{equation}
\varphi^{c}\equiv i\tau^{2}\varphi^{*}=\left(\begin{array}{c}
\varphi^{0*}\\
-\varphi^{-}
\end{array}\right)
\end{equation}
which transforms as a doublet, with the required hypercharge $\boldsymbol{y}_{\varphi^{c}}=-\frac{1}{2}$.
\end{doublespace}

\subsection*{1.5 Spontaneous Symmetry Breaking (SSB)}

\begin{doublespace}
Spontaneous symmetry breaking occurs because the lowest energy vacuum
state, represented by $\left|0\right\rangle $, does not respect gauge
symmetry and thereby induces effective masses for particles propagating
through it. We therefore consider the complex vector formed by: 
\begin{equation}
v=\left\langle 0\right|\varphi\left|0\right\rangle =\mathrm{const.}
\end{equation}
which has components that are vacuum expectation values (VEVs) of
the components of the complex scalar fields. We set $\, v=\mathrm{constant}$,
because any space or time dependence would increase the energy of
the solution, taking us away from the minimum. Similar terms for the
fermion $\,\left\langle 0\right|\mathcal{\psi}\left|0\right\rangle \,$
and boson $\,\left\langle 0\right|\mathcal{A}_{\mu}\left|0\right\rangle \,$
fields are zero because of Lorentz invariance. We determine $\, v\,$
by substituting $\,\varphi\rightarrow v=\left\langle 0\right|\varphi\left|0\right\rangle \,$
into the scalar potential $\, V\left(\varphi\rightarrow v\right)\,$
and minimizing $\, V\left(v\right)\,$ with respect to $\, v\,$.
The single complex Higgs doublet in the standard model can be re-written
in a Hermitian basis $\left(\varphi_{i}=\varphi_{i}^{\dagger},\; i=1..4\right)$
as: 
\begin{equation}
\varphi=\left(\begin{array}{c}
\varphi^{+}\\
\varphi^{0}
\end{array}\right)=\frac{1}{\sqrt{2}}\left(\begin{array}{c}
\varphi_{1}-i\varphi_{2}\\
\varphi_{3}-i\varphi_{4}
\end{array}\right)
\end{equation}

In the $\,\varphi_{i}\,$ basis, the Higgs potential becomes: 
\begin{equation}
V(\varphi)=\frac{1}{2}\mu^{2}\left(\sum_{i=1}^{4}\varphi_{i}^{2}\right)+\frac{1}{4}\lambda\left(\sum_{i=1}^{4}\varphi_{i}^{2}\right)^{2}
\end{equation}
Since this is $O_{4}$ invariant, we are allowed to choose an axis
in this 4d space such that $\,\langle0|\varphi_{i}|0\rangle=0,\;\ i=1,2,4\,$,
and $\,\langle0|\varphi_{3}|0\rangle=\nu$. In this basis the scalar
potential transforms as: 
\begin{equation}
V(\varphi)\rightarrow V(\nu)=\frac{1}{2}\mu^{2}\nu^{2}+\frac{1}{4}\lambda\nu^{\ensuremath{4}}
\end{equation}
Minimizing with respect to $\,\nu\,$, for the case when $\,\mu^{2}<0\,$,
we obtain: 
\begin{equation}
V'(\nu)=\nu(\mu^{2}+\lambda\nu^{2})=0
\end{equation}
therefore, the minimization condition is $\,\nu=\left(-\mu^{2}/\lambda\right)^{1/2}\,$
and the Higgs doublet gets replaced by its classical value: 
\begin{equation}
\varphi=\left(\begin{array}{c}
\varphi^{+}\\
\varphi^{0}
\end{array}\right)\rightarrow\, v=\frac{1}{\sqrt{2}}\left(\begin{array}{c}
0\\
\nu
\end{array}\right)
\end{equation}

We can quantize the theory by expanding around this minimum, $\,\varphi=\nu+\varphi'$.
We can go to a basis where the Goldstone bosons disappear by performing
a Kibble transformation \cite{Kibble1967} on the four Hermitian components
of $\,\varphi'\,$ , and employing the unitary gauge. 
\begin{equation}
\varphi\rightarrow\varphi'=e^{-i\sum\xi^{i}\tau^{i}}\varphi=\frac{1}{\sqrt{2}}\left(\begin{array}{c}
0\\
\nu+H
\end{array}\right)
\end{equation}
Here, $\, H\,$ is the Higgs scalar, the $\,\xi^{i}\,$ $\,\left(i=1,2,3\right)\,$
are three Hermitian fields, and $\,\tau^{i}\,$ are the Pauli matrices
. 

In the unitary gauge, the scalar covariant kinetic energy term takes
the form:
\begin{equation}
(\mathcal{D}_{\mu}\varphi)^{\dagger}\mathcal{D}^{\mu}\varphi=\frac{1}{2}(\begin{array}{cc}
0 & \nu\end{array})\left[\frac{g}{2}\tau^{i}W_{\mu}^{i}+\frac{g'}{2}\mathcal{B}_{\mu}\right]^{2}\left(\begin{array}{c}
0\\
\nu
\end{array}\right)+\mathrm{higgs}{\rm \; terms}
\end{equation}
\begin{equation}
\rightarrow M_{W}^{2}W^{+\mu}W_{\mu}^{-}+\frac{M_{Z}^{2}}{2}Z^{\mu}Z_{\mu}+\mathrm{higgs}{\rm \; terms}
\end{equation}
where the kinetic energy and gauge interaction terms of the physical
$\, H\,$ particle have been omitted. Therefore, SSB gives rise to
mass terms for the $W$ and $Z$ gauge bosons, in the form of the
following mixtures of gauge eigenstates : 
\begin{equation}
W_{\mu}^{\pm}=\frac{1}{\sqrt{2}}(W^{1}\mp iW^{2})
\end{equation}
\begin{equation}
Z_{\mu}=-\sin\theta_{W}\mathcal{B}_{\mu}+\cos\theta_{W}W^{3}
\end{equation}
After SSB, the massless photon field $\mathcal{A}_{\mu}$ appears
as well: 
\begin{equation}
\mathcal{A_{\mu}}=\cos\theta_{W}\mathcal{B}_{\mu}+\sin\theta_{W}W^{3}
\end{equation}
 In addition, we get mass terms for the charged weak gauge bosons:
\begin{equation}
M_{W^{\pm}}=\frac{g\nu}{2}
\end{equation}
 and the neutral weak gauge boson: 
\begin{equation}
M_{Z}=\sqrt{g^{2}+g^{\prime2}}\,\frac{\nu}{2}=\frac{M_{W}}{\cos\theta_{W}}
\end{equation}
The Goldstone boson has disappeared from the theory but has re-emerged
as the longitudinal degree of freedom of a massive vector particle.
The weak angle is defined by
\begin{equation}
\tan\theta_{W}\equiv g'/g.
\end{equation}
The weak scale is $\nu$ is given by 
\begin{equation}
\nu=2M_{W}/g\simeq\left(\sqrt{2}G_{F}\right)^{-1/2}\simeq246\,\mathrm{GeV}
\end{equation}

Similarly, electric charge appears through the relation, $\, g=e/\sin\theta_{W}\,$,
where $\, e\,$ is the electric charge of the positron. Thus we have
\begin{equation}
M_{W}=M_{Z}\cos\theta_{W}\thicksim\frac{\left(\pi\alpha/\sqrt{2}G_{F}\right)^{1/2}}{\sin\theta_{W}}
\end{equation}
 Taking account of the fact that $\sin^{2}\theta_{W}\sim0.23$ and
$\alpha\sim1/129$ , we arrive the leading order masses $M_{Z}=91\,\mathrm{GeV}$
and $M_{W}=80\,\mathrm{GeV}$.

After symmetry breaking the Higgs potential becomes: 
\begin{equation}
V(\varphi)=-\frac{\mu^{4}}{4\lambda}-\mu^{2}H^{2}+\lambda\nu H^{3}+\frac{\lambda}{4}H^{4}
\end{equation}
The Higgs mass term is given by: 
\begin{equation}
M_{H}=\sqrt{-2\mu^{2}}=\sqrt{2\lambda}\nu
\end{equation}
The quartic Higgs coupling $\lambda$ is unknown, so $M_{H}$ is not
predicted \emph{a priori} by the SM.

The Yukawa interaction in the unitary gauge becomes 
\begin{equation}
-\mathcal{L}_{{\rm Yukawa}}=\sum_{\psi}\left(\sum_{m,n=1}^{3}\bar{\psi}_{mL}Y_{mn}^{\left\{ u\right\} }\left(\frac{\nu+H}{\sqrt{2}}\right)\psi_{mR}+.h.c.\right)
\end{equation}
\begin{equation}
=\sum_{\psi}\bar{\psi}_{L}\left(M_{mn}^{\left\{ \psi\right\} }+h_{mn}^{\left\{ \psi\right\} }H\right)\psi_{R}
\end{equation}
where the sum over $\psi$ means write one term for each fermion.
\begin{equation}
\psi_{L}=\left(\begin{array}{c}
\psi_{1}\\
\psi_{2}\\
\psi_{3}
\end{array}\right)_{L}
\end{equation}

\end{doublespace}

$M_{mn}^{\left\{ \psi\right\} }$ is the fermion mass matrix for the
fermion $\psi$, 
\begin{equation}
M_{mn}^{\left\{ \psi\right\} }=Y_{mn}^{\left\{ \psi\right\} }\frac{\nu}{\sqrt{2}}
\end{equation}
 induced by spontaneous symmetry breaking, and
\begin{equation}
h_{mn}^{\left\{ \psi\right\} }=\frac{1}{\nu}M_{mn}^{\left\{ \psi\right\} }=\frac{g}{2}\frac{M_{mn}^{\left\{ \psi\right\} }}{M_{W}}
\end{equation}
 is the associated Yukawa coupling matrix for the fermion $\psi$.

\begin{doublespace}
In general the fermion mass matrix is not diagonal, Hermitian or symmetric.
To identify the physical particle content it is necessary to diagonalize
$M_{mn}^{\left\{ \psi\right\} }$ by separate unitary transformations
$\mathcal{A}_{L}$ and $\mathcal{A}_{R}$ on the left and right-handed
fermion fields. For example, if the fermion is an up-type quark then,
\begin{equation}
\mathcal{A}_{L}^{a\dagger}M^{b}\mathcal{A}_{R}^{c}=M_{D}^{u}=\left(\begin{array}{ccc}
m_{u} & 0 & 0\\
0 & m_{c} & 0\\
0 & 0 & m_{t}
\end{array}\right)
\end{equation}
is a diagonal matrix with eigenvalues equal to the physical masses
of the charge $\frac{2}{3}$ quarks. Of course, if $M_{mn}^{\psi}$
is Hermitian one can take $\mathcal{A}_{L}=\mathcal{A}_{R}$ without
consequence. Similarly, one diagonalizes the down quark and charged
lepton mass matrices by
\begin{equation}
\mathcal{A}_{L}^{d\dagger}M^{d}\mathcal{A}_{R}^{d}=M_{D}^{d}
\end{equation}
\begin{equation}
\mathcal{A}_{L}^{e\dagger}M^{e}\mathcal{A}_{L}^{e\dagger}=M_{D}^{e}.
\end{equation}
 In terms of these unitary matrices we can define mass eigenstate
fields
\begin{equation}
U_{L}=\mathcal{A}_{L}^{u\dagger}U_{L}^{0}=(U_{L}\ C_{L}\ T_{L})^{T}
\end{equation}
\begin{equation}
U_{R}=\mathcal{A}_{R}^{u\dagger}U_{R}^{0}=(U_{R}\ C_{R}\ T_{R})^{T}
\end{equation}
with analogous definitions for $D_{L,R}=\mathcal{A}_{L,R}^{d\dagger}D_{L,R}^{0}$,
and $E_{L,R}=\mathcal{A}_{L,R}^{e\dagger}E_{L,R}^{0}$. 

The Lagrangian for the SM after SSB is 
\[
\mathcal{L}=\mathcal{L}_{{\rm gauge}}+\mathcal{L}_{{\rm Higgs}}+\sum_{i}\bar{\Psi}_{i}\left(i\not{\!\partial}_{\mu}-m_{i}-\frac{m_{i}H}{\nu}\right)\Psi_{i}
\]
\begin{equation}
-\frac{g}{2\sqrt{2}}\left(\mathcal{J}_{W}^{\mu}W_{\mu}^{-}+\mathcal{J}_{W}^{\mu\dagger}W_{\mu}^{+}\right)-e\mathcal{J}_{Q}^{\mu}\mathcal{A}_{\mu}-\frac{g}{2\cos\theta_{W}}\mathcal{J}_{Z}^{\mu}\mathcal{Z}_{\mu}
\end{equation}
where the weak charged current is 
\begin{equation}
\mathcal{J}_{W}^{\mu\dagger}=\sum_{m=1}^{F}\left[\bar{\nu}_{m}\gamma^{\mu}(1-\gamma^{5})E_{m}+\bar{U}_{m}\gamma^{\mu}(1-\gamma^{5})D_{m}\right]
\end{equation}
\begin{equation}
=(\bar{\nu}_{e}\bar{\nu}_{\mu}\bar{\nu}_{\tau})\gamma^{\mu}(1-\gamma^{5})\left(\begin{array}{c}
e^{-}\\
\mu^{-}\\
\tau^{-}
\end{array}\right)+(\bar{u}\;\bar{c}\;\bar{t})\gamma^{\mu}(1-\gamma^{5})V_{\mathrm{CKM}}\left(\begin{array}{c}
d\\
s\\
b
\end{array}\right)
\end{equation}
and the quark flavor mixing CKM matrix is 
\begin{equation}
V_{CKM}=\left(\begin{array}{ccc}
V_{ud} & V_{us} & V_{ub}\\
V_{cd} & V_{cs} & V_{cb}\\
V_{td} & V_{td} & V_{td}
\end{array}\right).
\end{equation}
The electromagnetic current is given by 
\begin{equation}
\mathcal{J}_{Q}^{\mu}=\sum_{m=1}^{F}\left[\frac{2}{3}\bar{u}_{m}\gamma^{\mu}u_{m}-\frac{1}{3}\bar{d}_{m}\gamma^{\mu}d_{m}-\bar{e}_{m}\gamma^{\mu}e_{m}\right]
\end{equation}
and the Weak Neutral Current is given by 
\begin{equation}
\mathcal{J}_{Z}^{\mu}=\sum_{m}\left[\bar{u}_{mL}\gamma^{\mu}u_{mL}-\bar{d}_{mL}\gamma^{\mu}d_{mL}+\bar{\nu}_{mL}\gamma^{\mu}\nu_{mL}-\bar{e}_{mL}\gamma^{\mu}e_{mL}\right]-2\sin^{2}\theta_{W}\mathcal{J}_{Q}^{\mu}.
\end{equation}

In the limit $|Q^{2}|\ll M_{W}^{2}$, the momentum term in the $W$
propagator can be neglected, leading to an effective zero-range four
Fermi interaction: 
\begin{equation}
-\mathcal{L}_{{\rm eff}}^{cc}=\frac{G_{F}}{\sqrt{2}}\mathcal{J}_{W}^{\mu}\mathcal{J}_{W\mu}^{\dagger}
\end{equation}
where the Fermi constant is identified as 
\begin{equation}
\frac{G_{F}}{\sqrt{2}}\simeq\frac{g^{2}}{8M_{W}^{2}}=\frac{1}{2\nu^{2}}\,.
\end{equation}
Thus, the experimentally measured muon lifetime, $G_{F}=1.16639(2)\times10^{-5}\mathrm{GeV^{-2}}$,
implies that the weak interaction scale defined by the vacuum expectation
value (VEV) of the Higgs field, is given by: 
\begin{equation}
\nu=\sqrt{2}\langle0|\varphi|0\rangle\simeq246\,\mathrm{GeV.}
\end{equation}

\end{doublespace}

\begin{doublespace}

\subsection*{1.6 Problems with the SM}
\end{doublespace}

\begin{doublespace}
The mathematically consistent, renormalizable standard electroweak
model, is a field theory which is in agreement with all experimental
facts \cite{Nakamura2010}. It successfully predicted the existence
and the correct form of the weak neutral current, and the existence
and masses of the W and Z bosons. The SM successfully incorporates
the generalized Fermi theory with the prediction of the weak charged
current interactions. It also successfully incorporates quantum electrodynamics.
When combined with quantum chromodynamics and classical general relativity,
the predictions of the SM have been verified to a level of precision
which imply that it is certainly the approximately correct description
of nature down to at least $10^{-18}\:\mathrm{m}$. However, the SM
is too arbitrary to be truly fundamental, having some 21 free parameters. 

There is also the so called gauge problem, which is the fact that
there is no explanation for why only the electroweak sector is chiral.
Moreover, the standard model incorporates but does not explain charge
quantization; it provides no explanation for the reason that all particles
have charges which are multiples of $e/3$. Possible explanations
include: grand unified theories (GUTS), the existence of magnetic
monopoles, and constraints from the absence or cancellation of anomalies.

Perhaps the most fatal of all problems concerning the SM is the hierarchy
problem, which is essentially the question of why the Higgs mass is
so light relative to fundamental scales . In the standard model one
introduces, by hand, an elementary Higgs field into the theory to
generate masses for the $W$, $Z$, and fermions. For the model to
be consistent we need $\, m_{H}^{2}=O(m_{W}^{2})$. If $\, m_{H}$
is larger than $\, m_{W}$ by several orders of magnitude then we
arrive at a hierarchy problem. The tree-level, bare, Higgs mass receives
quadratically divergent corrections from one loop corrections, i.e.
$\, m_{H}^{2}=(m_{H}^{2})_{{\rm bare}}+O(\lambda,g^{2},h^{2})\Lambda^{2}$,
where $\Lambda$ is the energy scale where the effects of new physics
begins to make non-negligible contributions. We know \emph{ab initio
}that the SM breaks down at the scale where gravity becomes strong,
i.e the Planck scale, thus the extreme upper limit is $\,\Lambda=M_{P}=G_{N}^{-1/2}\sim10^{19}\,\mathrm{GeV}$. 

Naturalness suggests the unification of all forces at some higher
energy scale. This occurs in many grand unified theories (GUTs) at
scales on the order of $M_{X}\sim10^{16}\mathrm{GeV}$ and thus the
natural scale for $M_{H}$ would be $O(\Lambda)$, much larger than
the expected value obtained from the SM. In either case, in order
for the SM to make accurate predictions, as it does, there must be
a fine tuned, highly contrived cancellation between the bare value
and the correction, to more than 30 decimal places in the case of
gravity \cite{Lee1972}. Moreover, if the cutoff is provided by a
grand unified theory, we get an additional hierarchy problem at tree
level. The tree level couplings between the Higgs field and the extremely
heavy fields lead to the expectation that $M_{H}$ is equal to the
unification scale $M_{X}$, unless unnatural fine-tunings are done
to an embarrassingly high level of precision. 

Baryon $B$, and Lepton $L$, number are automatically global symmetries
of the SM, thus one may be naturally motivated to consider extensions
of the SM in which the individual symmetries of $B$ and $L$ , or
combinations there-of are gauged into locally symmetric extensions
of the SM gauge group. One reason for doing this, as we will see,
is that particular versions of combined $B$ and $L$ locally symmetric
extensions of the SM is their ability to incorporate small non-zero
neutrino masses in a natural way, and even posses viable DM candidates
without the need going to supersymmetric theories.

The most promising solution to this fine tuning problem is supersymmetry
(SUSY) which prevents large renormalization by providing a natural
mechanism for cancellations between the various Higgs mass corrections.
Thus in order to attack the shortcomings of the SM, as well as the
experimental shortcomings presented above, we consider SUSY extensions
of the SM as they provide a means to simultaneously resolve these
issues.

\clearpage{}
\end{doublespace}

\section*{Chapter 2}

\section*{Superspace and Supersymmetry}

\subsection*{2.1 Introduction and motivation}

\begin{doublespace}
We have seen that in the SM at tree-level the bare Higgs mass receives
quadratically divergent corrections from one loop processes: 
\begin{equation}
m_{H}^{2}=(m_{H}^{2})_{{\rm bare}}+O(\lambda,g^{2},h^{2})\Lambda^{2}\:.
\end{equation}

For example, corrections to the Higgs mass appear if the Higgs field
couples to a fermion $f$ with a term in the Lagrangian like the following
\begin{equation}
\mathcal{L}_{Hf}=-\lambda_{f}H\bar{f}f\:.
\end{equation}
This term is represented by the Feynman diagram in Figure 1(a), and
it yields the following correction to the Higgs mass 
\begin{equation}
\Delta m_{H}^{2}=-\frac{|\lambda_{f}|^{2}}{8\pi^{2}}\Lambda^{2}+\ldots\label{eq:62}
\end{equation}
Moreover, suppose there exists a heavy complex scalar particle $\, S\,$
with mass $\, m_{S}\,$ that couples to the Higgs with a Lagrangian
term like$\,-\lambda_{S}|H|^{2}|S|^{2}$. The Feynman diagram for
this process is shown in Figure 1(b). It gives rise to the following
correction to the Higgs mass: 
\begin{equation}
\Delta m_{H}^{2}=\frac{\lambda_{S}}{16\pi^{2}}\left[\Lambda^{2}-2m_{S}^{2}+{\rm ln}(\Lambda/m_{S})+\ldots\right]\label{eq:63}
\end{equation}

Comparing equations \eqref{eq:62} and \eqref{eq:63} we see that
if we consider a new symmetry relating fermions and bosons , then
the fine-tuning problems of the SM might be dealt with in a natural
way \cite{Martin1997}. Motivated along these lines, it is apparent
that if we consider supersymmetry, whereby the quarks and leptons
of the Standard Model each get accompanied by two complex scalars
with $\lambda_{S}=2|\lambda_{f}|^{2}$, then
\begin{equation}
\left(\Delta m_{H}^{2}\right)_{f}+\left(\Delta m_{H}^{2}\right)_{S}=-\frac{\lambda_{S}}{16\pi^{2}}\Lambda^{2}+\frac{\lambda_{S}}{16\pi^{2}}\Lambda^{2}-\frac{\lambda_{S}}{8\pi^{2}}m_{S}^{2}+\frac{\lambda_{S}}{16\pi^{2}}\left[{\rm ln}(\Lambda/m_{S})+\ldots\right]
\end{equation}
\begin{equation}
=-\frac{\lambda_{S}}{8\pi^{2}}m_{S}^{2}+\frac{\lambda_{S}}{16\pi^{2}}\left[{\rm ln}(\Lambda/m_{S})+\ldots\right]
\end{equation}
evidently the $\Lambda^{2}$ contributions of the loops in Figure
1 (a) and (b) cancel exactly, and we have at most logarithmic divergences
which can be summed with renormalization group procedures \cite{Dimopoulos1981,Dine1981,Witten1981}.
A viable theoretical framework that incorporates weakly-coupled Higgs
bosons is that of \textquotedblleft{}low energy\textquotedblright{}
or \textquotedblleft{}weak-scale\textquotedblright{} supersymmetry.
In this framework, supersymmetry is used to relate fermion and boson
masses and interaction strengths. Supersymmetry is a symmetry whereby
every fermionic/bosonic SM particle has an associated superpartner
which differs in spin by $1/2$.
\end{doublespace}

\noindent \begin{center}
\includegraphics[scale=0.3]{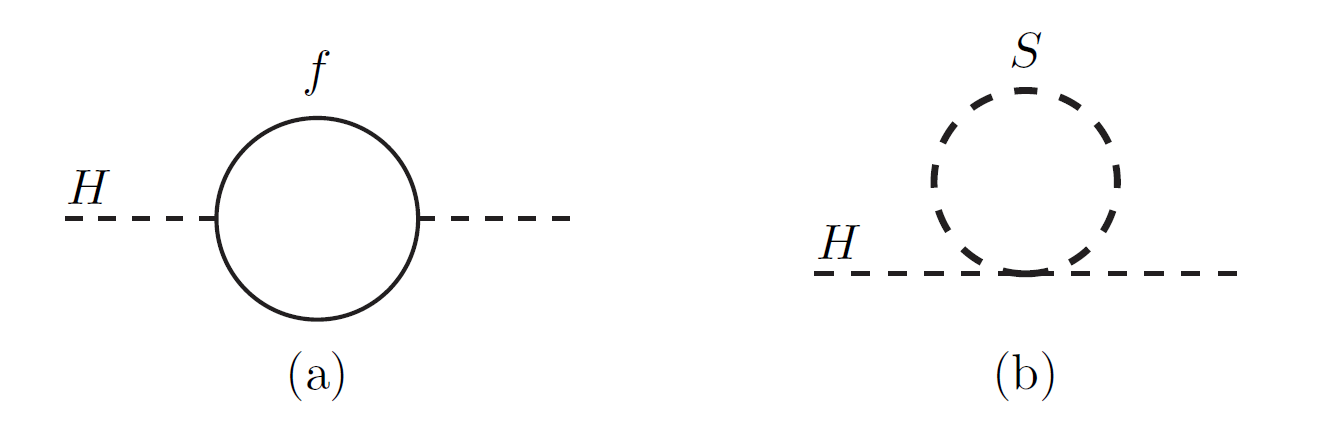}
\par\end{center}

\noindent \begin{center}
Figure 1: One-loop quantum corrections to the Higgs squared mass parameter
$m_{H}^{2}$, from (a) a Dirac fermion $f$ , and (b) a scalar $S$.
\par\end{center}

\subsection*{2.2 The SUSY Algebra and its Representations}

\begin{doublespace}
A supersymmetry transformation turns a bosonic state into a fermionic
state, and vice versa. The operator $Q$ that generates such transformations
must be an anti-commuting spinor, with 
\begin{equation}
Q|{\rm Boson}\rangle=|{\rm Fermion}\rangle\qquad\qquad Q|{\rm Fermion}\rangle=|{\rm Boson}\rangle
\end{equation}
In a 4-dimensional spacetime the minimal spinor is a Weyl spinor and
therefore the minimal supersymmetry has 4 supercharges. Supersymmetry
appeared for the first time in 1971 , as a phenomenological consideration.
The attempt to explain the neutrino as a Goldstone fermion associated
with the spontaneous breaking of a fermionic symmetry compelled Y.
Golfand and E. Likhtman to introduce the supersymmetric extension
of the Poincare algebra \cite{Golfand1971}.

In 1967 Coleman and Mandula proved a no-go theorem regarding the combination
of the newly discovered $SU\left(3\right)_{i}$ internal flavor symmetries
with $SU(2)_{spin}$ into $SU\left(6\right)$. Coleman and Mandula
proved that this combination did not respect S-matrix symmetries \cite{Coleman1967}.
They were only considering the special case where the generators commute.
If we also allow for an anti-commuting component, the Coleman Mandula
theorem no longer holds.

When we allow for fermionic generators, i.e. anti-commuting generators,
the theorem is no longer valid and we can extend the Poincare symmetry
of spacetime to include anti-commuting spinorial generators as was
shown by Haag, Lopusza\textasciiacute{}nski, and Sohnius in 1975 \cite{Haag1975}.
They proved that supersymmetry is the only additional symmetry of
the S-matrix allowed by inclusion of anti-commuting generators of
the Poincare algebra. This extension forms a graded-Lie algebra defined
by the usual commutation relations of the Poincare symmetry together
with the new anti-commutation relations, (where, $\alpha,\beta\in1,2$
and $\mu=0,1,2,3$) : 
\begin{equation}
\{Q_{\alpha}^{A},\bar{Q}_{\dot{\beta}}^{B}\}=2\sigma_{\alpha\dot{\beta}}^{\mu}\delta_{AB}\, P_{\mu}
\end{equation}
\begin{equation}
\{Q_{\alpha}^{A},Q_{\beta}^{B}\}=\epsilon_{\alpha\beta}X^{AB};\;\:\{\bar{Q}_{\dot{\alpha}}^{A},\bar{Q}_{\dot{\beta}}^{B}\}=\epsilon_{\dot{\alpha}\dot{\beta}}X^{AB}
\end{equation}
\begin{equation}
[Q_{\alpha}^{A},P_{\mu}]=[\bar{Q}_{\dot{\beta}}^{B},P_{\mu}]=0
\end{equation}
 Here $\, P_{\mu}\,$ is the generator of translations and the $X^{AB}s$
are anti-symmetric Lorentz scalars, known as the central charges.
The four dimensional, $\mathcal{N}=1$ SUSY algebra, which forms the
basis of the models we consider below, has no central charges since
$A,B$ can only take on one value and the $\, X^{AB}\,$ then comprise
a single anti-commuting Lorentz scalar, which therefore must vanish. 

The irreducible massive states in the 4 dimensional $\,\mathcal{N}=1\,$
SUSY theory obey an algebra which leads to some interesting and important
consequences. If we take the trace on both sides of the first equation
we obtain 
\begin{equation}
Q_{1}\bar{Q}_{\dot{1}}+\bar{Q}_{\dot{1}}Q_{1}+Q_{2}\bar{Q}_{\dot{2}}+\bar{Q}_{\dot{2}}Q_{2}=4P^{0}
\end{equation}
This is related to the Hamiltonian 
\begin{equation}
\mathcal{H}=P^{0}=\frac{1}{4}\left(Q_{1}\bar{Q}_{\dot{1}}+\bar{Q}_{\dot{1}}Q_{1}+Q_{2}\bar{Q}_{\dot{2}}+\bar{Q}_{\dot{2}}Q_{2}\right)
\end{equation}
Any state in the theory that is invariant under a symmetry is annihilated
by symmetry generators. In particular, if the ground state $|{\rm 0}\rangle$
is supersymmetric, it will be annihilated by SUSY generators, i.e.
$\, Q_{\alpha}|0\rangle=0\,$.

In general, for a non-zero energy eigenstate we have: 
\begin{equation}
\left\langle E\right|Q_{1}\bar{Q}_{\dot{1}}+\bar{Q}_{\dot{1}}Q_{1}+Q_{2}\bar{Q}_{\dot{2}}+\bar{Q}_{\dot{2}}Q_{2}|E\rangle=\left\langle E\right|4P^{0}|E\rangle=4E
\end{equation}
Because of the SUSY algebra, non-zero states must come in Fermi-Bose
pairs, i.e.
\begin{equation}
Q|B\rangle=\sqrt{E}|F\rangle,\; Q|F\rangle=\sqrt{E}|B\rangle
\end{equation}
In global SUSY theories, $E=0$, and $\, Q_{\alpha}|0\rangle=0\,$
, therefore $E$ is an order parameter for SUSY breaking. The ground
state of SUSY is dependent on the order parameter $\, E\,$ which
parametrizes the ground state of the SUSY theory in its broken $E>0,$
and unbroken $\, E=0$$\,$ phases. We, therefore, conclude that the
energy of a supersymmetric ground state must be zero. On the other
hand, if supersymmetry is spontaneously broken, the vacuum energy
is positive definite.
\end{doublespace}

\subsection*{2.3 Irreducible Representations of the SUSY Algebra}

\begin{doublespace}
There are two cases to consider, massless and massive representations.
We begin with the\emph{ first case, }and seek irreducible representations
of massive states, $M>0$, where $M=\mathrm{mass}$. All massive frames
are obtained from Wigner boosts and rotations applied to the rest
frame $\, P_{\mu}=(M,0,0,0)\,$ in which the velocity is zero. The
SUSY Algebra in this case is an algebra of $\,2N\,$ fermion creation
$\,\bar{Q}_{\dot{\beta}B}\,,$ and annihilation $\, Q_{\alpha}^{A}\,$
operators
\begin{equation}
\{Q_{\alpha}^{\: A},\bar{Q}_{\dot{\beta}B}\}=2\delta_{\alpha\dot{\beta}}\delta_{\: B}^{A}\, P_{\mu}
\end{equation}
\begin{equation}
\left\{ Q_{\alpha}^{A},Q_{\beta}^{B}\right\} =\left\{ \bar{Q}_{\dot{\alpha}A},\bar{Q}_{\dot{\beta}B}\right\} =0
\end{equation}
 (strictly speaking we must scale each $\, Q\,$ by $\,\frac{1}{\sqrt{2M}}\,$
to get the standard form of creation and annihilation operators familiar
from Quantum Mechanics). An irreducible representation of SUSY in
this case has dimension $\,2^{2N}.$

The \emph{second case} is massless, so $M=0$. The rest frame is given
by the four momentum $\, P_{\mu}=(E,0,0,E)\,$. Consider a particle
moving in the Z direction. $\,$
\begin{equation}
\{Q_{\alpha}^{A},\bar{Q}_{\dot{\beta}}^{B}\}=2\sigma_{\alpha\dot{\beta}}^{\mu}\delta_{AB}\, P_{\mu}=\delta_{AB}2E\left(\begin{array}{cc}
1 & 0\\
0 & 0
\end{array}\right)_{\alpha\dot{\beta}}
\end{equation}

For massless case we get $N$ creation and annihilation operators,
So the irreducible representation has dimension $\,2^{N}\,$ and this
is an irreducible representation of SUSY. However, if we include the
dimensionality of the CPT (equivalent to Lorentz symmetry) generators
along with those of SUSY, the general dimension is $\,2\cdot2^{N}=2^{N+1}\,$.
\end{doublespace}

\begin{doublespace}

\subsection*{2.4 Superspace and the Superfield Formalism}
\end{doublespace}

\begin{doublespace}
It would be very useful if we had a way to treat particles and their
superpartners as a single field (or superfield). Scalars and fermions
related by supersymmetry would simply correspond to different components
of a single superfield. To arrive at the desired superfield formalism
it is convenient to introduce the notion of superspace by extending
the 4 commuting spacetime coordinates $\, x_{\mu}\,$ to 4 commuting
and 4 anti-commuting coordinates $\,\{x_{\mu},\theta^{\alpha},\bar{\theta}_{\dot{\alpha}}\}\,$,
where $\,\bar{\theta}_{\dot{\alpha}}=(\theta^{\alpha})^{*}\,$. These
co-ordinates satisfy the following anti-commutation relations
\begin{equation}
\left\{ \theta_{\alpha},\bar{\theta}_{\dot{\beta}}\right\} =\left\{ \theta_{\alpha},\theta_{\beta}\right\} =\left\{ \bar{\theta}_{\dot{\alpha}},\bar{\theta}_{\dot{\beta}}\right\} =0
\end{equation}
We also introduce the integrals over superspace
\begin{equation}
\int d\theta=\int d\bar{\theta}=\int d\theta\,\bar{\theta}=\int d\bar{\theta}\,\theta=0\,,
\end{equation}
\begin{equation}
\int d\theta^{\alpha}\,\theta_{\beta}=\delta_{\beta}^{\alpha},\quad\quad\int d\bar{\theta}_{\dot{\alpha}}\bar{\theta}^{\dot{\beta}}=\delta_{\dot{\alpha}}^{\dot{\beta}}\,,
\end{equation}
\begin{equation}
\int d^{2}\theta\,\theta^{2}=\int d^{2}\bar{\theta}\,\bar{\theta}^{2}=1\,,\qquad\int d^{4}\theta\,\theta^{2}\bar{\theta}^{2}=1\,,
\end{equation}
where, 
\begin{equation}
d^{2}\theta\equiv-\frac{1}{4}\epsilon_{\alpha\beta}d\theta^{\alpha}d\theta^{\beta},
\end{equation}
\begin{equation}
d^{2}\bar{\theta}\equiv-\frac{1}{4}\epsilon_{\dot{\alpha}\dot{\beta}}d\theta^{\dot{\alpha}}d\theta^{\dot{\beta}},
\end{equation}
\begin{equation}
d^{4}\theta\equiv d^{2}\bar{\theta}d^{2}\theta
\end{equation}
The expansion of functions on superspace coordinates terminates at
order $\,\theta^{2}\bar{\theta}^{2}\,$. This property allows us to
express any supermultiplet as a single superfield which depends on
superspace coordinates. The Taylor expansion of the most general scalar
superfield is:
\begin{equation}
\Phi(\theta,\bar{\theta})=\phi+\theta\psi+\bar{\theta}\bar{\psi}+\bar{\theta}\bar{\sigma}^{\mu}\theta V_{\mu}+\theta^{2}F+\bar{\theta}^{2}\bar{F}+\ldots+\theta^{2}\bar{\theta}^{2}D
\end{equation}
The expansion above is a reducible representation of SUSY. To get
irreducible representations we consider chiral $\,\Phi\,$, and anti-chiral
$\,\Phi^{\dagger}\,$ superfields subject to the constraints 
\begin{equation}
\bar{D}_{\dot{\alpha}}\Phi=0\:,\quad\quad D_{\alpha}\Phi^{\dagger}=0\:,
\end{equation}
 where,
\begin{equation}
iQ_{\alpha}=D_{\alpha}=\frac{\partial}{\partial\theta^{\alpha}}-i\sigma_{\alpha\dot{\alpha}}^{\mu}\bar{\theta}^{\dot{\alpha}}\partial_{\mu}\:,
\end{equation}
\begin{equation}
i\bar{Q}_{\dot{\alpha}}=\bar{D}_{\dot{\alpha}}=-\frac{\partial}{\partial\bar{\theta}^{\dot{\alpha}}}+i\theta^{\alpha}\sigma_{\alpha\dot{\alpha}}^{\mu}\bar{\theta}^{\dot{\alpha}}\partial_{\mu}\:.
\end{equation}

If we introduce new variables $\, y^{\mu}=x^{\mu}+i\bar{\theta}\,\bar{\sigma}^{\mu}\theta\,$
and $\, y^{\mu\dagger}=x^{\mu}-i\bar{\theta}\,\bar{\sigma}^{\mu}\theta\,$,
then 
\begin{equation}
\bar{D}_{\dot{\alpha}}y^{\mu}=D_{\alpha}y^{\mu\dagger}=0.
\end{equation}
Thus we have chiral supermultiplets which are holomorphic functions
of $\theta$ and defined by 
\begin{equation}
\Phi(y^{\mu})=\varphi\left(y^{\mu}\right)+\sqrt{2}\theta\psi\left(y^{\mu}\right)+\theta^{2}F\left(y^{\mu}\right)\,.
\end{equation}
In addition, we have real vector supermultiplets which, in the Wess-Zumino
gauge, have the following form: 
\begin{equation}
V=-\theta\sigma^{\mu}\bar{\theta}V_{\mu}+i\left(\theta^{2}\bar{\theta}\bar{\lambda}-\bar{\theta}^{2}\theta\lambda\right)+\frac{1}{2}\theta^{2}\bar{\theta}^{2}D\,.
\end{equation}
The vector supermultiplet contains a vector field $\, V_{\mu}\,$
of mass dimension $1$, a gaugino $\,\lambda\,$ of mass dimension
$3/2$, and an auxiliary field $D$, of mass dimension 2, which will
be eliminated by solving the Euler-Lagrange (E-L) equations. 

In the following we will consider the MSSM and the $\, B-L\,$ extended
MSSM. These are both examples of $\,\mathcal{N}=1$ supersymmetric
theories. Generalizing the results above, the smallest irreducible
representations of $\,\mathcal{N}=1$ SUSY consists of the following
supermultiplets (in the Wess-Zumino gauge):
\end{doublespace}
\begin{itemize}
\begin{doublespace}
\item \noindent A vector superfield (VSF) for each gauge field. The physical
particle content of a VSF is one gauge boson and a Weyl fermion called
a \emph{gaugino}. The VSF's transform under the adjoint representation
of the gauge group.
\item \noindent A chiral superfield (CSF) for each matter field. The CSF
is composed of one spin-$\frac{1}{2}$ Weyl fermion and one spin-$0$
complex scalar. The CSF's can transform under any representation.
Since none of the matter fermions of the SM transform under the adjoint
of the gauge group we can not identify them with the gauginos. Thus
we have to introduce new fermionic SUSY partners to each SM gauge
boson.
\item \noindent A gravity supermultiplet containing the graviton (spin-$2$)
and its superpartner the gravitino (spin-$\frac{3}{2}$).\end{doublespace}

\end{itemize}

\subsection*{2.5 The MSSM}

\begin{doublespace}
The MSSM is a supersymmetric generalization of the Standard Model
gauge group $\,\mathcal{G_{SM}}$. It requires a color octet of vector
superfields $\mathcal{V}^{a}$, an additional weak triplet $\mathcal{V}^{i}$,
and a hypercharge singlet $\mathcal{V}$. These superfields contain
the appropriate spin-one gauge bosons and their spin-$\frac{1}{2}$
partners as displayed in table 3. In the MSSM the vector superfield
generalizations of the SM gauge fields, interact with the superfield
generalization of the quarks and leptons. These superfields are also
shown in Table 3. They are chiral superfields; they contain the spin-$\frac{1}{2}$
quarks and leptons, as well as their spin zero partners, the squarks
and sleptons. The supersymmetric extensions of Higgs bosons are also
shown in Table 3. They include two complex Higgs doublets $\left(H_{u},H_{d}\right)$,
as well as their spin-$\frac{1}{2}$ partners, the two Higgsinos.
We introduce the extra fields because in SUSY theories, two (or more)
Higgs doublets are required for the Higgsino anomalies to cancel among
themselves, and they are also required to reproduce all the SM Yukawa
couplings.
\end{doublespace}

\begin{doublespace}
\noindent \begin{center}
\begin{tabular}{|c|c|c|c|c|c|c|}
\hline 
CSF & $SU\left(3\right)$ & $SU\left(2\right)$ & $U\left(1\right)$ & $B$ & $L$ & Particles\tabularnewline
\hline 
\hline 
$L_{i}$ & $1$ & $2$ & $-\frac{1}{2}$ & $0$ & $1$ & leptons $\left(\nu,e\right)$ and sleptons $\left(\tilde{\nu},\tilde{e}\right)$\tabularnewline
\hline 
$E_{i}^{c}$ & $1$ & $1$ & $1$ & $0$ & $-1$ & electron $e^{c}$ and selectron $\tilde{e}^{c}$\tabularnewline
\hline 
$Q_{i}$ & $3$ & $2$ & $+\frac{1}{6}$ & $\frac{1}{3}$ & $0$ & quarks $\left(u,d\right)$ and squarks $\left(\tilde{u},\tilde{d}\right)$\tabularnewline
\hline 
$U_{i}^{c}$ & $\bar{3}$ & $1$ & $-\frac{2}{3}$ & $-\frac{1}{3}$ & $0$ & quarks $u^{c}$ and squarks $\tilde{u}^{c}$\tabularnewline
\hline 
$D_{i}^{c}$ & $\bar{3}$ & $1$ & $\frac{1}{3}$ & $-\frac{1}{3}$ & $0$ & quarks $d^{c}$ and squarks $\tilde{d}^{c}$\tabularnewline
\hline 
$H_{u}$ & $1$ & $2$ & $\frac{1}{2}$ & $0$ & $0$ & Higgs $h_{u}$ and Higgsinos $\tilde{h}_{u}$\tabularnewline
\hline 
$H_{d}$ & $1$ & $2$ & $-\frac{1}{2}$ & $0$ & $0$ & Higgs $h_{d}$ and Higgsinos $\tilde{h}_{d}$\tabularnewline
\hline 
\hline 
VSF & $SU\left(3\right)$ & $SU\left(2\right)$ & $U\left(1\right)$ & $B$ & $L$ & Particles\tabularnewline
\hline 
$\mathcal{V}^{a}$ & 8 & 1 & 0 & 0 & 0 & gluons $g$ and gluinos $\tilde{g}$\tabularnewline
\hline 
$\mathcal{V}^{i}$ & 1 & 3 & 0 & 0 & 0 & $W$'s and winos $\tilde{W}$\tabularnewline
\hline 
$\mathcal{V}$ & 1 & 1 & 0 & 0 & 0 & $B$ and bino $\tilde{B}$\tabularnewline
\hline 
\end{tabular}
\par\end{center}
\end{doublespace}

\begin{singlespace}
\noindent \begin{center}
Table 3: The Particle contents of the MSSM $\left(i=1,2,3\right)$
\par\end{center}
\end{singlespace}

\begin{doublespace}
Once the particle content is fixed one can try to write down the most
general renormalizable Lagrangian for this $\mathcal{\, N}=1$ supersymmetric
$SU(3)\otimes SU(2)\otimes U(1)$ theory. It is well known from the
structure of $\,\mathcal{N}=1$ SUSY gauge theories that the Lagrangian
is completely fixed by gauge invariance and by supersymmetry, except
for the choice of the superpotential, whose most general form contains
all possible gauge invariant operators of dimension 3 or less. The
following superpotential contains all such operators. 

\begin{equation}
\mathcal{W}=Y_{u}^{ij}\, Q^{i}\, H_{u}\,\bar{U}^{j}+Y_{d}^{ij}Q^{i}H_{d}\bar{D}^{j}+Y_{e}^{ij}L^{i}H_{d}\bar{E}^{j}+\mu H_{d}H_{u}+\mathcal{W}_{\slashed{\mathrm{R}_{p}}}
\end{equation}
where the baryon and lepton number violating contributions are contained
in:
\begin{equation}
\mathcal{W}_{\slashed{\mathrm{R}_{p}}}=a_{1}^{ijk}Q^{i}L^{j}\bar{D}^{k}+a_{2}^{ijk}L^{i}L^{j}\bar{E}^{k}+a_{3}^{i}L^{i}H_{u}+a_{4}^{ijk}\bar{D}^{i}\bar{D}^{j}\bar{U}^{k}
\end{equation}

The terms on the first line, with the exception of $\,\mathcal{W}_{\slashed{\mathrm{R}_{p}}}$,
correspond to the SUSY generalization of the ordinary Yukawa interactions
of the SM plus an additional $\mu$-term breaking the Peccei-Quinn
degeneracy of the two Higgs doublet model. The terms on the second
line are generally allowed since they are gauge invariant, but lead
to lepton and baryon number violating interactions. Baryon and lepton
number non-conservation is at odds with the situation in the SM where
the most general renormalizable gauge invariant Lagrangian automatically
conserved baryon and lepton number. To proceed we must impose additional
symmetries which forbid $B$ and $L$ violating interactions that
are dangerous phenomenologically. 

The easiest way to achieve this is to introduce R-parity and require
that it be conserved. The R-parity of a field is given by
\begin{equation}
R=\left(-1\right)^{3B+L+2S}
\end{equation}
 where $B$ is the baryon number, $L$ the lepton number, and $S$
the spin of a given particle associated with a particular field. One
could also forbid the appearance of the $B$ and $L$ breaking terms
by imposing other, different symmetry requirements. For example a
$Z_{2}$ subgroup of $B\otimes L$ known as matter parity, 
\begin{equation}
P=\left(-1\right)^{3\left(B-L\right)}
\end{equation}
could achieve this goal as well. Actually matter parity and R-Parity
are equivalent for any vertex that conserves angular momentum \cite{Chung2005}.
The important point is that once lepton and baryon number violating
terms are absent, R-parity will necessarily be a symmetry of the Lagrangian
regardless of its ultimate status as primary or accidental. 

Another strong motivation for the introduction of R-Parity is that
it provides a dark matter candidate, namely the Lightest Supersymmetric
Particle (LSP). One consequence of R-parity conservation is that super-partners
may only be produced in pairs, implying that the LSP is stable if
R-parity is exactly conserved.
\end{doublespace}

\subsection*{2.6 Supersymmetry Breaking}

\begin{doublespace}
If SUSY were exact then the masses of all the sparticles would be
identical to their SM counterparts. For example, there would be a
particle, the selectron, which had the same mass as the electron,
and we would certainly have seen this particle by now. Therefore SUSY,
if it exists, must exists as a broken symmetry in nature. There are
two possibilities for SUSY breaking, explicit SUSY breaking and spontaneous
SUSY breaking. A model whose Lagrangian density is not invariant under
supersymmetric transformations explicitly breaks SUSY. On the other
hand, a\emph{ }spontaneous SUSY breaking model is one whose Lagrangian
density is invariant under supersymmetry, but whose vacuum state is
not. Spontaneous SUSY breaking is favored from a theoretical point
of view because we would always prefer our Lagrangian densities to
be SUSY invariant in a SUSY theory , however this is not a viable
option in the MSSM \cite{Csaki1996}. 

Our only remaining option then is to introduce explicit SUSY breaking
terms in order to break SUSY. However these terms should not come
at price of sacrificing the solution to the hierarchy problem. Such
terms are called soft SUSY breaking terms, and those are the terms
that do not reintroduce quadratic divergences into the theory. The
soft supersymmetry breaking Lagrangian is defined to include all allowed
terms that do not introduce quadratic divergences in the theory: all
gauge invariant and Lorentz invariant terms of dimension two and three.The
complete set of possible soft SUSY breaking parameters was elucidated
by K. Inoue, A. Kakuto, H. Komatsu and S. Takeshita \cite{Inoue1982}
, and the classic proof provided by Girardello and Grisaru \cite{Girardello1982}.
The $\mathcal{L}_{soft}$ terms are of the following types, where
the summation convention is implied throughout:
\end{doublespace}
\begin{itemize}
\item Soft tri-linear scalar interactions: 
\begin{equation}
\frac{1}{3!}\mathcal{A}_{ijk}\phi_{i}\phi_{j}\phi_{k}+h.c.
\end{equation}

\item Soft bi-linear scalar interactions: 
\begin{equation}
\frac{1}{2}\mathcal{B}_{ij}\phi_{i}\phi_{j}+h.c.
\end{equation}

\item Soft scalar mass-squares: 
\begin{equation}
m_{ij}^{2}\phi_{i}^{\dagger}\phi_{j}
\end{equation}

\item Soft gaugino masses (where, $a$ is a group label): 
\begin{equation}
\frac{1}{2}\mathcal{M}_{a}\lambda^{a}\lambda^{a}+h.c.
\end{equation}

\end{itemize}
\begin{doublespace}
The basic idea behind the soft breaking terms is that there exists
a sector of physics that breaks SUSY spontaneously. This sector resides
at energy scales much higher than the weak scale. SUSY breaking is
then communicated in some way (either through gauge interactions or
through gravity) to the MSSM fields and as a result the soft breaking
terms appear. 

A common implementation proceeds to break SUSY spontaneously in a
hidden sector which is decoupled from the SM particles in the visible
sector, except through supergravity which will mediate the SUSY breaking
terms to the visible sector. The minimal supergravity mediation mechanism
generates universal soft breaking terms for the visible sector fields
at the Planck scale. Thus one has to think of the MSSM as an effective
theory, valid below a certain scale (of new physics), and the soft
breaking terms will parametrize our ignorance of the details of the
physics of the SUSY breaking sector.
\end{doublespace}

\subsection*{2.7 mSUGRA and the CMSSM}

\begin{doublespace}
The most general soft SUSY breaking terms introduce 105 new parameters
into our model, this is definitely not a good thing because of known
constraints on FCNC's and CP violation. The most minimal approach
to mediating supersymmetry breaking between hidden and visible sectors
is through supergravity interactions. Generically one should expect
that the most general interactions consistent with the symmetries
of both hidden and visible sector will be generated in an effective
theory with Planck suppressed couplings. The Gravity mediation of
SUSY breaking from the hidden sector is flavor blind and so it drastically
reduces the number of free parameters by justifying the assumption
of diagonal SUSY breaking mass matrices. In the mSUGRA mediation scenario,
and we obtain the so-called constrained MSSM or CMSSM. In this case
we assume unification of the following parameters at the GUT scale
$\mathcal{M}_{GUT}$ \cite{Iso2011}:
\end{doublespace}
\begin{itemize}
\item Universal gaugino masses (at the the GUT Scale):
\begin{equation}
\mathcal{M}_{3}=\mathcal{M}_{2}=\mathcal{M}_{1}=\mathcal{M}_{1/2}
\end{equation}

\item Universal scalar masses(sfermions and Higgs) masses(at the the GUT
Scale):
\begin{equation}
\mathcal{M}_{Q}^{2}=\mathcal{M}_{L}^{2}=\mathcal{M}_{U^{c}}^{2}=\mathcal{M}_{D^{c}}^{2}=\mathcal{M}_{E^{c}}^{2}=\mathcal{M}_{H_{u}}^{2}=\mathcal{M}_{H_{d}}^{2}=m_{0}^{2}
\end{equation}

\item Universal tri-linear couplings (at the the GUT Scale):
\begin{equation}
A_{u}=A_{d}=A_{L}=A_{0}
\end{equation}

\end{itemize}
This reduces the number of free parameters from 124 to just 5 \cite{Baer2006}:
\begin{equation}
\tan\beta,\quad M_{1/2},\quad m_{0},\quad A_{0}\quad\mathrm{sign\left(\mu\right)}
\end{equation}
Where, $\tan\beta=\left|\left\langle H_{u}\right\rangle \right|/\left|\left\langle H_{d}\right\rangle \right|$,
and $\mu$ is the co-efficient of the Peccei-Quinn term in $\mathcal{W}_{MSSM}$.

\subsection*{2.8 The MSSM Higgs Sector}

\begin{doublespace}
The Higgs scalar potential is given by
\begin{equation}
V=V_{D}+V_{F}+V_{soft}
\end{equation}
The term $V_{D}$ represents the D-term potential which is obtained
from 
\begin{equation}
V_{D}=\sum_{A}\frac{1}{2}D^{A}D^{A}
\end{equation}
 where
\begin{equation}
D^{A}\equiv-g_{A}\phi_{i}^{*}T_{ij}^{A}\phi_{j}
\end{equation}
The $U(1)_{Y}$ contribution to $D$ term is
\begin{equation}
D^{1}=-\frac{g'}{2}\left(\left|H_{u}\right|^{2}-\left|H_{d}\right|^{2}\right)
\end{equation}
and the $SU(2)$ contribution to the $D$ term is (where $T^{a}=\frac{\tau^{a}}{2}$):
\[
D^{a}=-\frac{g}{2}\left(H_{d}^{i*}\tau_{ij}^{a}H_{d}^{j}+H_{u}^{i*}\tau_{ij}^{a}H_{u}^{j}\right)
\]
 Thus the D-terms contribute the following to the scalar potential
\begin{equation}
V_{D}=\frac{g^{\prime2}}{8}\left(\left|H_{d}\right|^{2}-\left|H_{u}\right|^{2}\right)^{2}+\frac{g^{2}}{8}\left(H_{d}^{i*}\tau_{ij}^{a}H_{d}^{j}+H_{u}^{i*}\tau_{ij}^{a}H_{u}^{j}\right)^{2}
\end{equation}
Using the $SU(2)$ identity
\begin{equation}
\tau_{ij}^{a}\tau_{kl}^{a}=2\delta_{il}\delta_{jk}-\delta_{ij}\delta_{kl}
\end{equation}
we may write this as: 
\begin{equation}
V_{D}=\frac{g^{2}}{8}\left[4\left|H_{u}^{*}\cdot H_{d}\right|^{2}-2\left(H_{u}^{*}\cdot H_{d}\right)\left(H_{d}^{*}\cdot H_{d}\right)+\left(\left|H_{u}\right|^{2}+\left|H_{d}\right|^{2}\right)\right]+\frac{g^{\prime2}}{8}\left(\left|H_{d}\right|^{2}-\left|H_{u}\right|^{2}\right)^{2}
\end{equation}
which, upon recalling the definitions 
\begin{equation}
H_{u}=\left(\begin{array}{c}
H_{u}^{+}\\
H_{u}^{0}
\end{array}\right),\quad H_{d}=\left(\begin{array}{c}
H_{d}^{0}\\
H_{d}^{-}
\end{array}\right)
\end{equation}
can be written in component form as: 
\begin{equation}
V_{D}(H_{u},H_{d})=\frac{g^{2}+g^{\prime2}}{8}\left(|H_{u}^{0}|^{2}+|H_{u}^{+}|^{2}-|H_{d}^{0}|^{2}-|H_{d}^{-}|^{2}\right)^{2}+\frac{g^{2}}{2}\left|H_{u}^{+}H_{d}^{0*}+H_{u}^{0}H_{d}^{-*}\right|^{2}
\end{equation}
The $F$-term is given by $V_{F}=\sum_{i}\left|F_{i}\right|^{2}$,
where $F_{i}^{*}=\frac{\partial\mathcal{W}}{\partial\phi_{i}}$, and
the $\,\phi_{i}\,$ are the scalar components of the chiral superfields
in the MSSM superpotential. The $F$-term component of the scalar
potential is :
\begin{equation}
V_{F}=\mu^{2}\left(|H_{u}^{0}|^{2}+|H_{u}^{+}|^{2}+|H_{d}^{0}|^{2}+|H_{d}^{-}|^{2}\right)
\end{equation}
The soft SUSY breaking terms in the potential are given by ($\tilde{m}_{u}^{2}$
and $\tilde{m}_{d}^{2}$ are soft Higgs masses):
\end{doublespace}

\begin{singlespace}
\[
V_{soft}=\tilde{m}_{u}^{2}\left(|H_{u}^{0}|^{2}+|H_{u}^{+}|^{2}\right)+\tilde{m}_{d}^{2}\left(H_{d}^{0}|^{2}+|H_{d}^{-}|^{2}\right)+
\]
\begin{equation}
+B\left(H_{u}^{+}H_{d}^{-}-H_{u}^{0}H_{d}^{0}\right)+B\left(H_{u}^{+*}H_{d}^{-*}-H_{u}^{0*}H_{d}^{0*}\right)
\end{equation}

\end{singlespace}

\begin{doublespace}
EWSB requires that the Higgs potential is bounded from below and has
a minimum at non-vanishing VEVs. The Higgs mass matrix will satisfy
these conditions if

\begin{equation}
|B|^{2}>(\tilde{m}_{u}^{2}+|\mu|^{2})(\tilde{m}_{d}^{2}+|\mu|^{2})
\end{equation}
and,
\begin{equation}
2\mu^{2}+\tilde{m}_{u}^{2}+\tilde{m}_{d}^{2}>2|B|.
\end{equation}

\end{doublespace}

\subsection*{2.9 The Renormalization Group Equations}

\begin{doublespace}
The MSSM is free of quadratic divergences, RGE evolution modifies
relations between superpartner masses and may lead to radiative EWSB.
The dominant effect arises from the Higgs interactions with the third
generation where $\left(\mathrm{where}\;\ensuremath{t=\frac{1}{16\pi^{2}}\log\frac{M_{GUT}^{2}}{\Lambda^{2}}}\right)$
:
\begin{equation}
\frac{d}{dt}\left(\begin{array}{c}
\tilde{m}_{u}^{2}\\
\tilde{m}_{\tilde{t}}^{2}\\
\tilde{m}_{Q_{3}}^{2}
\end{array}\right)=Y_{t}\left(\begin{array}{ccc}
3 & 3 & 3\\
2 & 2 & 2\\
1 & 1 & 1
\end{array}\right)-A_{t}^{2}\left(\begin{array}{c}
3\\
2\\
1
\end{array}\right)
\end{equation}

We can see that $H_{u}$ receives the largest negative contribution
and once its mass is driven negative electroweak symmetry is broken.
Thus we can see that the radiative corrections due to the top Yukawa
coupling want to reverse the sign of the soft breaking mass parameter
of the up-type Higgs, which is enough to satisfy the conditions for
electroweak breaking at the weak scale. Appropriate choices of the
input parameters $\, M_{1/2},\, m_{0},\, A_{0}\,$ and $\,\lambda_{t}\,$
will drive the soft breaking mass parameter of the up-type Higgs negative
which will result in the breaking of electroweak symmetry. This mechanism
is called radiative electroweak symmetry breaking.

By requiring that the parameters in the Higgs potential lead to experimentally
observed $Z$ and $W$ mass we obtain relations between soft parameters
which must be satisfied at the weak scale: 
\begin{equation}
\mu^{2}=\frac{\tilde{m}_{u}^{2}-\tilde{m}_{d}^{2}\tan\beta}{\tan^{2}\beta-1}-\frac{1}{2}M_{Z}^{2}
\end{equation}
\begin{equation}
B=\frac{1}{2}\left(\tilde{m}_{u}^{2}+\tilde{m}_{d}^{2}+2\mu^{2}\right)\sin2\beta
\end{equation}
 where $\tan\beta=\left\langle H_{u}\right\rangle /\left\langle H_{d}\right\rangle $.

From this expression we see that naturalness requires that both $\,\mu\,$
and $\, B\,$ are electroweak scale parameters. However, the $\mu$-term
is a supersymmetric term in the Lagrangian and could take any value
between EWSB and Planck scales. This leads to the so-called $\mu$-problem.
Only models where the $\mu$-term arises as a result of SUSY breaking
are expected to avoid fine-tuning. Even then, the absence of fine-tuning
is not guaranteed. 

The MSSM solves the hierarchy problem of the SM but it still cannot
explain the origin of neutrino masses. The B-L extension of the MSSM
allows for a natural implementation of the see-saw mechanism, the
well known natural way to obtain small non-zero neutrino masses. \clearpage{}
\end{doublespace}

\section*{\noindent Chapter 3}

\section*{\noindent The B-L Extension of The MSSM}

\subsection*{3.1 Introduction}

\begin{doublespace}
It is well-known that the relevant scale for neutrino mass generation
through the seesaw mechanism is given by the $\, B-L\,$ scale. The
global $\, B\lyxmathsym{\textminus}L\,$ symmetry present in the SM
is an accidental anomaly free symmetry, and once it is gauged, the
$\, U(1)_{B-L}^{3}$ and $\, U\left(1\right)_{B-L}\,$ anomalies must
be canceled. The $B\lyxmathsym{\textminus}L$ anomaly cancellation
conditions can be satisfied by introducing three generations of right-handed
neutrinos, which are singlets under the SM.

The particle content of the Supersymmetric $\, B-L\,$ extended Standard
Model includes the following fields in addition to those of the MSSM: 
\end{doublespace}
\begin{itemize}
\item Three chiral right-handed neutrino superfields $\left\{ \nu_{1}^{c},\,\nu_{2}^{c},\,\nu_{3}^{c}\right\} $. 
\item The $\, Z_{B-L}\,$ vector superfield necessary to gauge the $\, U(1)_{B-L}\,$
symmetry. 
\item Two chiral standard model singlet Higgs superfields $\,\left(X,\,\bar{X}\right)\,$
with $\, B-L\,$ charges $\,\left(-2,\,2\right)\,$ respectively. 
\end{itemize}
\begin{doublespace}
As is the case in the MSSM, the introduction of a second Higgs singlet
$\bar{X}$ is necessary in order to cancel the $\, U(1)_{B-L}\,$
anomalies produced by the fermionic member of the first Higgs superfield
$\, X\,$. The charges for quark and lepton superfields are assigned
in the usual way. We present the particle contents of all supermultiplets
of the $\, B-L\,$ extended MSSM below in table 4:
\end{doublespace}

\noindent \begin{center}
\begin{tabular}{|c||c||c||c||c|}
\hline 
\noalign{\vskip\doublerulesep}
 & $SU\left(3\right)_{c}$ & $SU\left(2\right)_{L}$ & $U\left(1\right)_{Y}$ & $U\left(1\right)_{B-L}$\tabularnewline[\doublerulesep]
\hline 
\hline 
\noalign{\vskip\doublerulesep}
$Q_{i}$ & $3$ & $2$ & $+\frac{1}{6}$ & $+\frac{1}{3}$\tabularnewline[\doublerulesep]
\hline 
\hline 
\noalign{\vskip\doublerulesep}
$U_{i}^{c}$ & $\bar{3}$ & $1$ & $-\frac{2}{3}$ & $-\frac{1}{3}$\tabularnewline[\doublerulesep]
\hline 
\hline 
\noalign{\vskip\doublerulesep}
$D_{i}^{c}$ & $\bar{3}$ & $1$ & $\frac{1}{3}$ & $-\frac{1}{3}$\tabularnewline[\doublerulesep]
\hline 
\hline 
\noalign{\vskip\doublerulesep}
$L_{i}$ & $1$ & $2$ & $-\frac{1}{2}$ & $-1$\tabularnewline[\doublerulesep]
\hline 
\hline 
\noalign{\vskip\doublerulesep}
$E_{i}^{c}$ & $1$ & $1$ & $1$ & $+1$\tabularnewline[\doublerulesep]
\hline 
\hline 
\noalign{\vskip\doublerulesep}
$\nu_{1}^{c}$ & $1$ & $1$ & $0$ & $+1$\tabularnewline[\doublerulesep]
\hline 
\hline 
\noalign{\vskip\doublerulesep}
$\nu_{2}^{c}$ & $1$ & $1$ & $0$ & $+1$\tabularnewline[\doublerulesep]
\hline 
\hline 
\noalign{\vskip\doublerulesep}
$\nu_{3}^{c}$ & $1$ & $1$ & $0$ & $+1$\tabularnewline[\doublerulesep]
\hline 
\hline 
\noalign{\vskip\doublerulesep}
$H_{u}$ & $1$ & $2$ & $+\frac{1}{2}$ & $0$\tabularnewline[\doublerulesep]
\hline 
\hline 
\noalign{\vskip\doublerulesep}
$H_{d}$ & $1$ & $2$ & $-\frac{1}{2}$ & $0$\tabularnewline[\doublerulesep]
\hline 
\hline 
\noalign{\vskip\doublerulesep}
$X$ & $1$ & $1$ & $0$ & $-2$\tabularnewline[\doublerulesep]
\hline 
\hline 
\noalign{\vskip\doublerulesep}
$\bar{X}$ & $1$ & $1$ & $0$ & $+2$\tabularnewline[\doublerulesep]
\hline 
\end{tabular}
\par\end{center}

\begin{doublespace}
\noindent \begin{center}
Table 4: Particle Contents of $B-L$ extended MSSM
\par\end{center}
\end{doublespace}

\begin{doublespace}
The superpotential for the model is \cite{FileviezPerez2011}:

\begin{equation}
\mathcal{W}=\mathcal{W}_{MSSM}+\mathcal{W}_{B-L}
\end{equation}
\begin{equation}
\mathcal{W}_{MSSM}=Y_{u}\, Q\, H_{u}\, u^{c}+Y_{d}\, Q\, H_{d}\, d^{c}+Y_{e}\, L\, H_{d}\, e^{c}+\mu\, H_{u}\, H_{d}
\end{equation}
\begin{equation}
\mathcal{W}_{B-L}=Y_{\nu}L\, H_{u}\nu^{c}+f\,\nu^{c}\nu^{c}X-\mu_{X}\, X\,\bar{X}
\end{equation}

The soft supersymmetric breaking Lagrangian is 
\begin{equation}
-\mathcal{L}_{soft}=a_{\nu}L\, H_{u}\nu^{c}-a_{X}\nu^{c}\nu^{c}X-b_{X}\, X\,\bar{X}+\frac{1}{2}M_{BL}B'B'+\ h.c.
\end{equation}
\begin{equation}
+m_{X}^{2}\, X^{2}+m_{\bar{X}}^{2}\,\bar{X}^{2}+m_{\nu^{c}}^{2}\left(\nu^{c}\right)^{2}
\end{equation}

Where $B'$ is the $B-L$ gaugino. Spontaneous $B-L$ violation requires
either the VEV of $\, X\,$, $\,\bar{X}\,$ or $\,\nu^{c}\,$ to be
nonzero, however R-Parity conservation depends on the VEV of $\nu^{c}$:
$\left<\nu^{c}\right>=0\,$ corresponds to $R$-parity conservation
while $\,\left<\nu^{c}\right>\neq0\,$ yields spontaneous $R$-parity
violation \cite{FileviezPerez2011}. 

In order to investigate the values of these VEV's we need the minimization
conditions which are derived from the full potential, (we use the
parametrization: $\left(\left<X\right>,\left<\bar{X}\right>,\left<\nu^{c}\right>\right)=1/\sqrt{2}\left(x,\bar{x},n\right)$),
which is:
\begin{equation}
\left\langle V\right\rangle =\left\langle V_{f}\right\rangle +\left\langle V_{D}\right\rangle +\left\langle V_{soft}\right\rangle 
\end{equation}
where,
\begin{equation}
\left\langle V_{f}\right\rangle =\frac{1}{4}f^{2}n^{4}+f^{2}n^{2}x^{2}+\frac{1}{2}\mu_{X}\left(x^{2}+\bar{x}^{2}\right)-\frac{1}{\sqrt{2}}f\,\mu_{x}n^{2}\bar{x}
\end{equation}
\begin{equation}
\left\langle V_{D}\right\rangle =\frac{1}{32}g_{BL}^{2}\left(2\bar{x}^{2}-2x^{2}+n^{2}\right)^{2}
\end{equation}
\begin{equation}
\left\langle V_{soft}\right\rangle =-\frac{1}{\sqrt{2}}\, a_{X}\, n^{2}x-b_{X}\, x\,\bar{x}+\frac{1}{2}m_{X}^{2}\, x^{2}+\frac{1}{2}m_{\bar{X}}^{2}\,\bar{x}^{2}+\frac{1}{2}m_{\bar{\nu}^{c}}n^{2}
\end{equation}

There are two mechanisms capable of inducing spontaneous $\, B-L\,$
violation, which are what we, and the authors in \cite{FileviezPerez2011}
have classified as: 
\end{doublespace}
\begin{itemize}
\item Case 1: $\,\left(n=0;\, x\neq0,\,\bar{x}\neq0\,\right)$ is the $R$-parity
conserving case. 
\item Case 2: $\,\left(x\neq0,\,\bar{x}\neq0,\, n\neq0\,\right)$ is the
$R$-parity violating case. 
\end{itemize}
\begin{doublespace}
\noindent There exists a third case:
\end{doublespace}
\begin{itemize}
\item \noindent Case 3: $\,\left(n\neq0;\, x=0,\,\bar{x}=0\,\right)$ ,
\end{itemize}
\begin{doublespace}
\noindent which cannot exist due to the linear term for $\, x\,$
in $V_{soft}$, and the linear term for $\,\bar{x}\,$ in $V_{F}$
which always induce a VEV for these fields. 
\end{doublespace}

\subsection*{3.2 The R-Parity Conserving Case}

\begin{doublespace}
The minimization conditions for $x$ and $\bar{x}$ are very similar
in form to those of $v_{u}$ and $v_{d}$ in the MSSM 
\begin{equation}
\frac{1}{2}M_{Z'}^{2}=-|\mu_{X}|^{2}+\frac{m_{X}^{2}\tan^{2}z-m_{\bar{X}}^{2}}{1-\tan^{2}z}\label{eq:130}
\end{equation}
where $\,\tan z\equiv x/\bar{x}\,$ and $\, M_{Z'}^{2}\equiv g_{BL}^{2}\left(x^{2}+\bar{x}^{2}\right)\,$,
which is the mass for the $\, Z'\,$ boson associated with broken
$\, B-L$. In this case it is useful to examine the limit where $\, x\gg\bar{x}\,$,
with $\, m_{X}^{2}<0\,$ and $\, m_{\bar{X}}^{2}>0\,$, in which equation
\ref{eq:130} reduces to: 
\begin{equation}
\frac{1}{2}M_{Z'}^{2}=-|\mu_{X}|^{2}-m_{X}^{2}
\end{equation}
so that, 
\begin{equation}
g_{BL}^{2}\left(x^{2}+\bar{x}^{2}\right)=-2\left(|\mu_{X}|^{2}+m_{X}^{2}\right)\label{eq:132}
\end{equation}
We see that for $B-L$ breaking to occur we require, in addition to
$m_{X}^{2}<0,$ that the magnitudes obey the constraint $\,-m_{X}^{2}>|\mu_{X}|^{2}\,$
.

This model holds promise for explaining neutrino masses. Replacing
$X$ by its VEV in the term $\, f\nu^{c}\nu^{c}X\,$ in the superpotential
leads to the heavy Majorana mass term for the right-handed neutrinos
and ultimately to the Type I seesaw mechanism for neutrino masses:
\begin{equation}
m_{\nu}=v_{u}^{2}\, Y_{\nu}^{T}\left(f\, x\right)^{-1}Y_{\nu}
\end{equation}
As can be seen from equation \ref{eq:132}, $\, x\,$ is of order
the SUSY mass scale or about a TeV. Moreover, given that the mass
of the right-handed neutrinos are of order TeV, scenarios which yield
realistic neutrino masses require, $Y_{\nu}\sim10^{-6-7}$. 
\end{doublespace}

\subsection*{3.3 The R-Parity Violating Case}

\begin{doublespace}
The evaluation of the minimization conditions in this case is illuminating
in the limit where $\, g_{BL}^{2}\,\ll\,1\,$, and $\, n\,\gg\left\{ x,\,\bar{x},\, a_{X}\right\} $
: 
\begin{equation}
n^{2}=\frac{-m_{\tilde{\nu}^{c}}^{2}\Lambda_{\bar{X}}^{2}}{f^{2}\, m_{\bar{X}}^{2}+\frac{1}{8}\, g_{BL}^{2}\Lambda_{\bar{X}}^{2}}
\end{equation}
\begin{equation}
\bar{x}=\frac{\left(-m_{\tilde{\nu}^{c}}^{2}\right)\, f\,\mu_{X}}{\sqrt{2}\left(f^{2}\, m_{\bar{X}}^{2}\ +\,\frac{1}{8}\, g_{BL}^{2}\,\Lambda_{\bar{X}}^{2}\right)}
\end{equation}
\begin{equation}
x=\frac{\left(-m_{\tilde{\nu}^{c}}^{2}\right)\left[a_{X}\Lambda_{\bar{X}}^{2}+f\ b_{X}\ \mu_{X}\right]}{\left(2\, f^{2}-\frac{1}{4}g_{BL}^{2}\right)\left(-m_{\tilde{\nu}^{c}}^{2}\right)\Lambda_{\bar{X}}^{2}+f^{2}m_{\bar{X}}^{2}\Lambda_{X}^{2}+\frac{1}{8}g_{BL}^{2}\Lambda_{\bar{X}}^{2}\Lambda_{X}^{2}}
\end{equation}

where, $\,\Lambda_{X}^{2}\equiv\mu_{X}^{2}+m_{X}^{2}\,$ and $\,\Lambda_{\bar{X}}^{2}\equiv\mu_{X}^{2}+m_{\bar{X}}^{2}\,$. 

The $Z'$ mass in the $R$-parity violating case is given by 
\begin{equation}
M_{Z'}^{2}=\frac{1}{4}\left(n^{2}+4x^{2}+4\bar{x}^{2}\right)
\end{equation}
We can see from these equations that spontaneous $B-L$ symmetry breaking
in the $R$-parity violating case only requires $\, m_{\nu^{c}}^{2}<0\,$.
Moreover, there is no introduction of a new $\,\mu\,$ problem so
that $\,\mu_{X}\,$ can be larger than the TeV scale. The $\,\mu\to\infty\,$
serves as a decoupling limit since $\, x,\,\bar{x}\to0$ and $\, n^{2}\to-8m_{\nu^{c}}/g_{BL}^{2}\,$
as in the minimal model.
\end{doublespace}

\begin{doublespace}

\subsection*{3.4 The RGE's and RSBM analysis}
\end{doublespace}

\begin{doublespace}
We will investigate $B-L$ parity conservation/violation in these
different parameter subspaces by evolving the RGEs down from the GUT
scale to the TeV scale. We use the gravitational mediation of SUSY
breaking in our analysis and here we will adopt the mSUGRA ansatz
with the following boundary conditions at the GUT scale: 
\begin{equation}
m_{X}^{2}=m_{\bar{X}}^{2}=m_{\nu_{i}^{c}}^{2}=m_{MSSM}^{2}=m_{0}^{2}
\end{equation}
 The boudary mass term $\, m_{0}\,$ is the universal scalar mass
at the GUT scale, and $\, m_{MSSM}^{2}\,$ indicates the relevant
set of MSSM parameters. The trilinear couplings also unify at the
GUT scale, as do the gauginos as shown below. In this case $\, Y_{MSSM}\,$
is the universal Yukawa coupling at the GUT scale. 
\begin{equation}
A_{X}=f\, A_{0};\quad A_{\nu}=Y_{\nu}A_{0};\quad A_{MSSM}=Y_{MSSM}A_{0}
\end{equation}
\begin{equation}
M_{BL}=M_{MSSM}=M_{1/2}
\end{equation}
Where $\, M_{1/2}\,$ is the universal gaugino mass.

We utilize the following renormalization group equations (RGEs), subjecting
them to the boundary conditions listed above.

The RGEs are given by \cite{Martin1994} (assuming a flavor diagonal
basis $f_{1}=f_{2}=f_{3}\,$) where $i=1,2,3$:

\begin{equation}
16\pi^{2}\,\frac{dg_{BL}}{dt}=\,9\, g_{BL}^{3}
\end{equation}
\begin{equation}
16\pi^{2}\,\frac{dM_{BL}}{dt}=18\, g_{BL}^{2}M_{BL}
\end{equation}
\begin{equation}
16\pi^{2}\,\frac{df_{i}}{dt}=f_{3}\left(8\, f_{i}^{2}+2\,\mathrm{Tr\,}f^{2}-\frac{9}{2}\, g_{BL}^{2}\right)
\end{equation}
\begin{equation}
16\pi^{2}\frac{da_{X_{i}}}{dt}=f_{X}\left(16\, f_{i}a_{X_{i}}+4\,\mathrm{Tr}\,\left(f\, a_{X}\right)-9\, g_{BL}^{2}M_{BL}\right)+a_{X_{i}}\left(8\, f_{i}^{2}+2\,\mathrm{Tr}\, f^{2}-\frac{9}{2}\, g_{BL}^{2}\right)
\end{equation}

The Running of the soft-masses is given by (for $\, i=1,2,3\,$):

\begin{equation}
16\pi^{2}\,\frac{dm_{\bar{X}}^{2}}{dt}=-12\, g_{BL}^{2}M_{BL}^{2}
\end{equation}
\begin{equation}
16\pi^{2}\,\frac{dm_{X}^{2}}{dt}=4\,\mathrm{Tr}\, f^{2}\, m_{X}^{2}+8\,\mathrm{Tr}\,\left(f^{2}m_{\nu^{c}}^{2}\right)\,+\,4\,\mathrm{Tr}\, a_{X}^{2}-12\, g_{BL}^{2}M_{BL}^{2}
\end{equation}
\begin{equation}
16\pi^{2}\,\frac{dm_{\nu_{i}^{c}}^{2}}{dt}=8\, f_{i}^{2}\left(m_{X}^{2}\,+\,2\, m_{\nu_{i}^{c}}^{2}\right)+\,8\, a_{X_{i}}^{2}-3\, g_{BL}^{2}M_{BL}^{2}
\end{equation}
Radiative symmetry breaking requires one of the soft masses to run
negative.
\end{doublespace}

\begin{doublespace}

\subsection*{3.5 Results}
\end{doublespace}

\begin{doublespace}
We have selected representative solutions for the different solution
subspaces. We chose specific values for the GUT scale couplings below,
however there is nothing inherently special about these numbers \emph{per
se}. We simply pick them to illustrate the observation of the existence
of three different solution subspaces. One could change the boundary
conditions to obtain different results for each solution space in
order to examine the details of any particular situation. Our goal
below is to present a single example of each of the three possible
scenarios, with boundary conditions characteristic of the solution
subspace in question.
\end{doublespace}
\begin{itemize}
\begin{doublespace}
\item \noindent (Case 1) For the $R$-parity conserving case, we have the
following boundary conditions at the GUT scale $M_{GUT}=3\cdot10^{16}\,\mathrm{GeV}$.
We let $\mu_{X}=2000\,\mathrm{GeV}$ in all cases.
\begin{equation}
f_{1}=0.855,\: f_{2}=0.935,\: f_{3}=1.095,\: g_{BL}^{2}=0.53,\: M_{1/2}=500,\: m_{0}=5000,\: A_{0}=0
\end{equation}
 We let the tri-linear couplings $\, A_{0}\,$ vanish as they have
very little effect on the overall running of the parameters. \end{doublespace}

\end{itemize}
\begin{doublespace}
\begin{minipage}[t]{1\columnwidth}%
\begin{doublespace}
\includegraphics[scale=0.4]{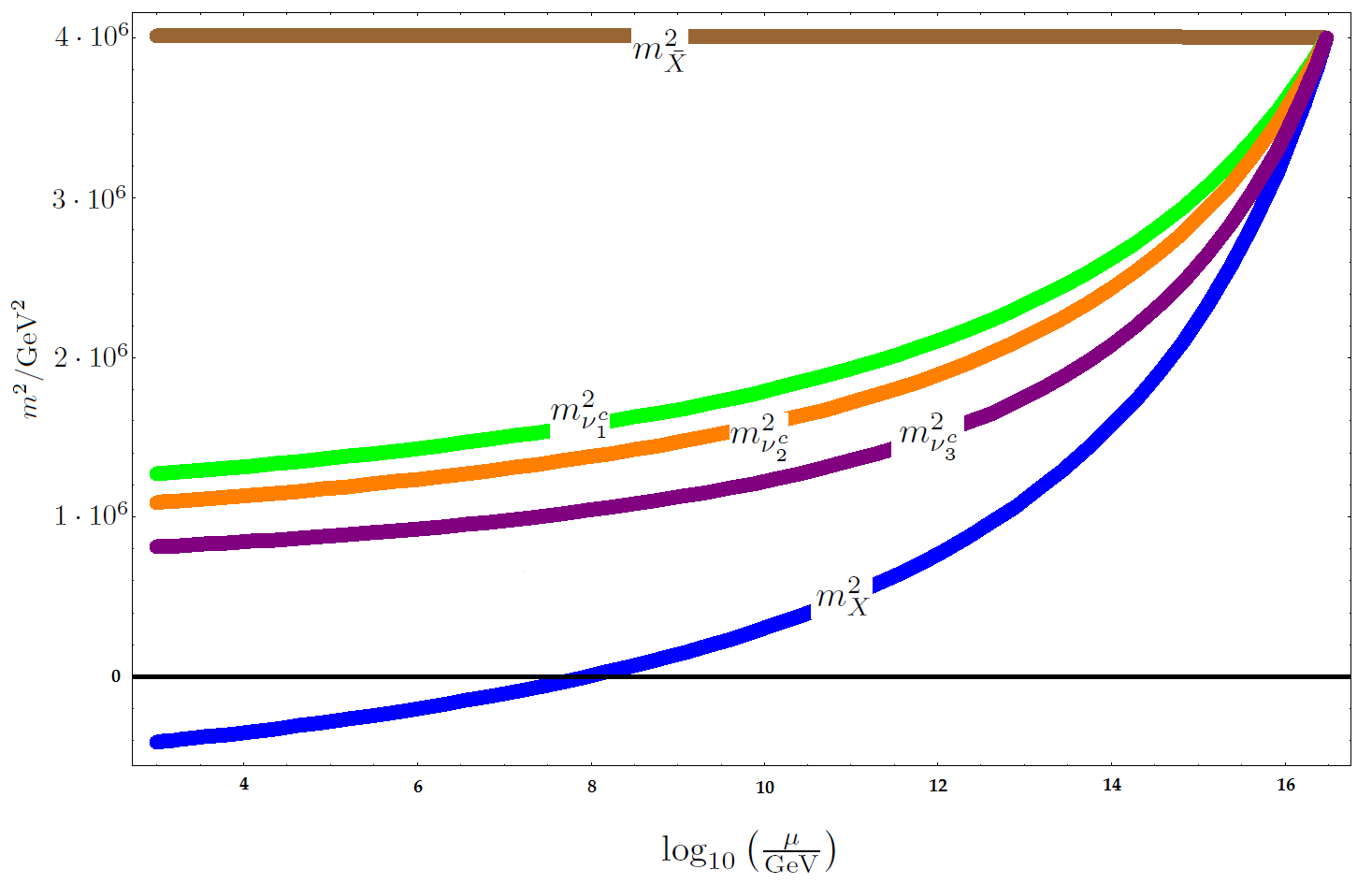}
\end{doublespace}

\begin{singlespace}
\noindent \begin{center}
Figure 2: RGE running of the scalar masses in\\
$R$-Parity conserving case (case 1) of the $B-L$ model.
\par\end{center}\end{singlespace}
\end{minipage}\vspace*{0.25in}

Examining figure 2, we see that the mass-squared of the $B-L=-2$
Higgs becomes negative, breaking the $B-L$ symmetry but preserving
$R$-parity. In this case the neutralino, the MSSM LSP, is still a
viable dark matter candidate since it is forbidden to decay through
$R$-parity conserving channels.
\end{doublespace}
\begin{itemize}
\begin{doublespace}
\item (Case 2) For the $R$-Parity Violating Case, we have the following
boundary conditions at the GUT scale $\, M_{GUT}=3\cdot10^{16}\,\mathrm{GeV}$.
\begin{equation}
f_{1}=f_{2}=0.4,\: f_{3}=3,\: g_{BL}^{2}=0.53,\: M_{1/2}=500,\: m_{0}=2000,\: A_{0}=0
\end{equation}
 We let the tri-linear couplings $A_{0}$ vanish, as they have very
little effect on the overall running of the parameters.\end{doublespace}

\end{itemize}
\begin{doublespace}
\begin{minipage}[t]{1\columnwidth}%
\begin{doublespace}
\includegraphics[scale=0.43]{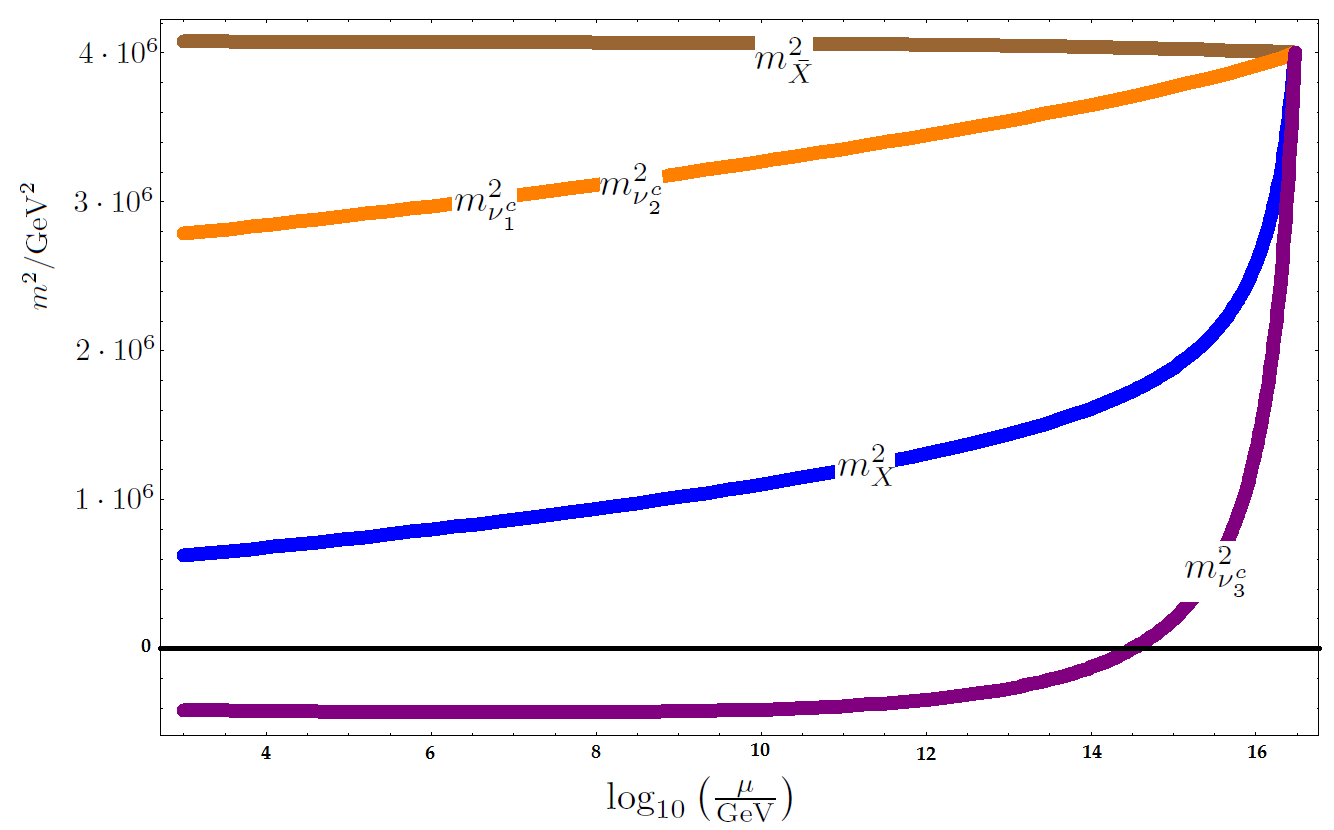}
\end{doublespace}

\begin{singlespace}
\noindent \begin{center}
Figure 3: RGE running of the scalar masses in \\
$R$-Parity violating case (case 2) of the $B-L$ model.
\par\end{center}\end{singlespace}
\end{minipage}\vspace*{0.25in}

Examining figure 3, we see that in this case the 3rd $\,\left(i=3\right)\,$
right handed neutrino, $\,\nu_{3}^{c}\,$ runs negative breaking $\, B-L\,$,
and simultaneously violating $R$-parity. 
\end{doublespace}
\begin{itemize}
\begin{doublespace}
\item (Case 3) This case does not violate $R$-parity or $\, B-L\,$. To
obtain a representative solution of this type let us take the following
boundary conditions at the GUT scale $M_{GUT}=3\cdot10^{16}\,\mathrm{GeV}$.
\begin{equation}
f_{1}=f_{2}=1.25,\: f_{3}=0.01,\: g_{BL}^{2}=0.53,\: M_{1/2}=500,\: m_{0}=2000,\: A_{0}=0
\end{equation}
We let the tri-linear couplings $A_{0}$ vanish as they have very
little effect on the overall running of the parameters. The situation
with large Yukawa couplings for the first and second right handed
neutrinos relative to the third leaves $\, B-L\,$ unbroken. \end{doublespace}

\end{itemize}
\begin{doublespace}
\begin{minipage}[c]{1\columnwidth}%
\begin{doublespace}
\includegraphics[scale=0.42]{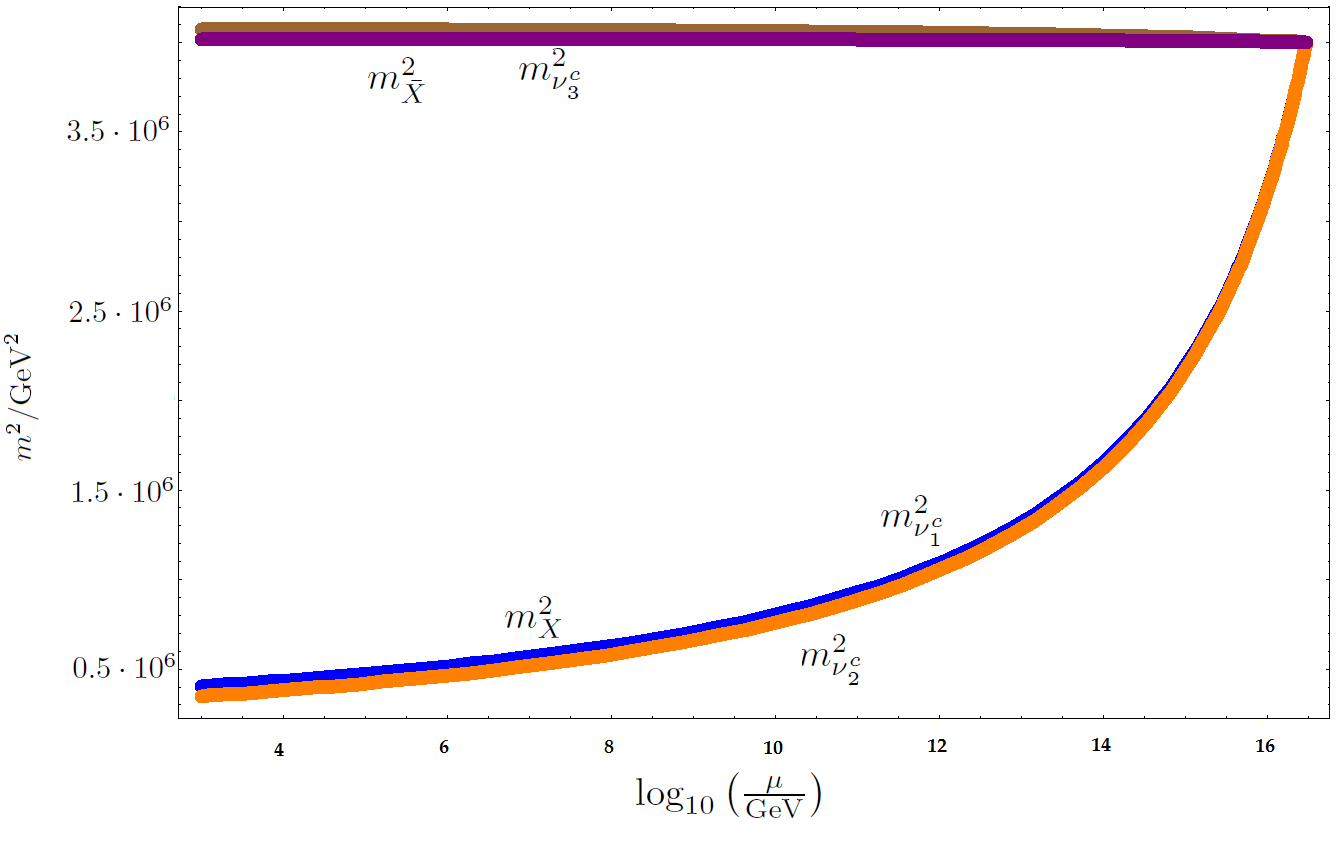}
\end{doublespace}

\begin{singlespace}
\noindent \begin{center}
Figure 4: RGE running of the scalar masses in \\
the dormant case (case 3) of the $B-L$ model.
\par\end{center}\end{singlespace}
\end{minipage}\vspace*{0.25in}

As can be seen from figure 4, with these boundary conditions none of the mass-squareds become tachyonic,
producing a phenomenologically dormant situation, never the less we include it here for
completeness.
\end{doublespace}

\subsection*{3.6 Additional $\mathbb{Z}_{2}$ parity and DM candidates in $R$-parity violating case.}

\begin{doublespace}
We have examined the $B-L$ extended MSSM and found that there
is a region of parameter space associated with this model that violates
$R$-parity even when we utilize mSUGRA mediation which protects the MSSM. To this end we postulate a new $\mathbb{Z}_{2}$ parity under which $\nu_{1}$ is $\mathbb{Z}_{2}$ odd, and everything else is $\mathbb{Z}_{2}$ even. Thus $\nu_{1}$ becomes a potential dark matter candidate, in addition to the gravitino; we display the
particle contents below with this additional $\mathbb{Z}_{2}$ parity. 
\end{doublespace}

\noindent \begin{center}
\begin{tabular}{|c||c||c||c||c||c|}
\hline 
\noalign{\vskip\doublerulesep}
 & $SU\left(3\right)_{c}$ & $SU\left(2\right)_{L}$ & $U\left(1\right)_{Y}$ & $\mathrm{Z_{2}}$ & $U\left(1\right)_{B-L}$\tabularnewline[\doublerulesep]
\hline 
\hline 
\noalign{\vskip\doublerulesep}
$Q_{i}$ & $3$ & $2$ & $+\frac{1}{6}$ & $+$ & $+\frac{1}{3}$\tabularnewline[\doublerulesep]
\hline 
\hline 
\noalign{\vskip\doublerulesep}
$U_{i}^{c}$ & $\bar{3}$ & $1$ & $-\frac{2}{3}$ & $+$ & $-\frac{1}{3}$\tabularnewline[\doublerulesep]
\hline 
\hline 
\noalign{\vskip\doublerulesep}
$D_{i}^{c}$ & $\bar{3}$ & $1$ & $\frac{1}{3}$ & $+$ & $-\frac{1}{3}$\tabularnewline[\doublerulesep]
\hline 
\hline 
\noalign{\vskip\doublerulesep}
$L_{i}$ & $1$ & $2$ & $-\frac{1}{2}$ & $+$ & $-1$\tabularnewline[\doublerulesep]
\hline 
\hline 
\noalign{\vskip\doublerulesep}
$E_{i}^{c}$ & $1$ & $1$ & $1$ & $+$ & $+1$\tabularnewline[\doublerulesep]
\hline 
\hline 
\noalign{\vskip\doublerulesep}
$\nu_{1}^{c}$ & $1$ & $1$ & $0$ & $-$ & $+1$\tabularnewline[\doublerulesep]
\hline 
\hline 
\noalign{\vskip\doublerulesep}
$\nu_{2}^{c}$ & $1$ & $1$ & $0$ & $+$ & $+1$\tabularnewline[\doublerulesep]
\hline 
\hline 
\noalign{\vskip\doublerulesep}
$\nu_{3}^{c}$ & $1$ & $1$ & $0$ & $+$ & $+1$\tabularnewline[\doublerulesep]
\hline 
\hline 
\noalign{\vskip\doublerulesep}
$H_{u}$ & $1$ & $2$ & $+\frac{1}{2}$ & $+$ & $0$\tabularnewline[\doublerulesep]
\hline 
\hline 
\noalign{\vskip\doublerulesep}
$H_{d}$ & $1$ & $2$ & $-\frac{1}{2}$ & $+$ & $0$\tabularnewline[\doublerulesep]
\hline 
\hline 
\noalign{\vskip\doublerulesep}
$X$ & $1$ & $1$ & $0$ & $+$ & $-2$\tabularnewline[\doublerulesep]
\hline 
\hline 
\noalign{\vskip\doublerulesep}
$\bar{X}$ & $1$ & $1$ & $0$ & $+$ & $+2$\tabularnewline[\doublerulesep]
\hline 
\end{tabular}
\par\end{center}

\vspace{0.3in.}
\begin{doublespace}

We plan to investigate the phenomenology of this case in a forthcoming paper.
\end{doublespace}

\clearpage{}

\bibliographystyle{plain}
\bibliography{thesis5.bib.bak}

\end{document}